\documentclass[draftcls,a4paper,11pt,onecolumn,oneside,journal]{IEEEtran}
\usepackage[utf8]{inputenc}
\usepackage[T1]{fontenc}
\usepackage{graphicx}
\usepackage{longtable}
\usepackage{wrapfig}
\usepackage{rotating}
\usepackage[normalem]{ulem}
\usepackage{amsmath}
\usepackage{textcomp}
\usepackage{amssymb}
\usepackage{capt-of}
\usepackage[bookmarks=false]{hyperref}
\usepackage{cite}
\usepackage{tabularx}
\usepackage{multirow}
\usepackage[table]{xcolor}
\usepackage{acronym}
\usepackage{multicol}
\usepackage{setspace}

\hypersetup{
 hidelinks,
 pdfauthor={Carlos H. M. de Lima},
 pdftitle={6G White Paper on Localization and Sensing},
 pdfkeywords={},
 pdfsubject={},
 pdfcreator={Emacs 26.3 (Org mode 9.1.9)}, 
 pdflang={English}}

\begin{document}

\title{6G White Paper on Localization and Sensing}

\author{\IEEEauthorblockN{Andre Bourdoux\( ^1 \), Andre Noll Barreto\( ^2 \), Barend van Liempd\( ^1 \), Carlos de Lima\( ^3 \), Davide Dardari\( ^4 \), Didier Belot\( ^5 \), Elana-Simona Lohan\( ^6 \), Gonzalo Seco-Granados\( ^7 \), Hadi Sarieddeen\( ^8 \), Henk Wymeersch\( ^9 \), Jaakko Suutala\( ^3 \), Jani Saloranta\( ^3 \), Maxime Guillaud\( ^{10} \), Minna Isomursu\( ^3 \), Mikko Valkama\( ^6 \), Muhammad Reza Kahar Aziz\( ^{11} \), Rafael Berkvens\( ^1 \), Tachporn Sanguanpuak\( ^3 \), Tommy Svensson\( ^9 \), Yang Miao\( ^{12} \)} \\
\begin{flushleft}
  \line(1,0){250} \\
  \IEEEauthorblockA{\( ^1 \)IMEC, Belgium \\
    \( ^2 \)Barkhausen Institut, Germany \\
    \( ^3 \)University of Oulu, Finland \\
    \( ^4 \)University of Bologna, Italy \\
    \( ^5 \)CEA-LETI, France \\
    \( ^6 \)Tampere University, Finland \\
    \( ^7 \)Universitat Autonoma de Barcelona, Spain \\
    \( ^8 \)King Abdullah University of Science and Technology (KAUST), Saudi Arabia \\
    \( ^9 \)Chalmers University of Technology, Sweden \\
    \( ^{10} \)Huawei Technologies Paris, France \\
    \( ^{11} \)Institut Teknologi Sumatera, Indonesia \\
    \( ^{12} \)University of Twente, Netherlands}
\end{flushleft}

}

\maketitle

\newpage

\begin{abstract}

This white paper explores future localization and sensing opportunities for beyond \ac{5G} wireless communication systems by identifying key technology enablers and discussing their underlying challenges, implementation issues, and identifying potential solutions. In addition, we present exciting new opportunities for localization and sensing applications, which will disrupt traditional design principles and revolutionize the way we live, interact with our environment, and do business. In contrast to \ac{5G} and earlier generations, localization and sensing will be built-in from the outset to both cope with specific applications and use cases, and to support flexible and seamless connectivity.
 
Following the trend initiated in the \ac{5G} \ac{NR} systems, \ac{6G} will continue to develop towards even higher frequency ranges, wider bandwidths, and massive antenna arrays. In turn, this will enable sensing solutions with very fine range, Doppler and angular resolutions, as well as localization to cm-level degree of accuracy. Moreover, new materials, device types, and reconfigurable surfaces will allow network operators to reshape and control the electromagnetic response of the environment. At the same time, machine learning and artificial intelligence will leverage the unprecedented availability of data and computing resources to tackle the biggest and hardest problems in wireless communication systems.
 
\ac{6G} systems will be truly intelligent wireless systems that will not only provide ubiquitous communication but also empower high accuracy localization and high-resolution sensing services. They will become the catalyst for this revolution by bringing about a unique new set of features and service capabilities, where localization and sensing will coexist with communication, continuously sharing the available resources in time, frequency and space. Applications such as THz imaging and spectroscopy have the potential to provide continuous, real-time physiological information via dynamic, non-invasive, contactless measurements for future digital health technologies. \ac{6G} \ac{SLAM} methods will not only enable advanced \ac{XR} applications but also enhance the navigation of autonomous objects such as vehicles and drones. In convergent \ac{6G} radar and communication systems, both passive and active radars will simultaneously use and share information, to provide a rich and accurate virtual image of the environment. In \ac{6G}, intelligent context-aware networks will be capable of exploiting localization and sensing information to optimize deployment, operation, and energy usage with no or limited human intervention. 

This white paper concludes by highlighting foundational research challenges, as well as implications and opportunities related to privacy, security, and trust. Addressing these challenges will undoubtedly require an inter-disciplinary and concerted effort from the research community.
\end{abstract}

\newpage


\tableofcontents

\newpage

\section{Introduction}
\label{sec:intro}

The \acf{5G} \acf{NR} is scheduled for worldwide release by \(2020\), thus, by now, its development cycle has reached initial maturity and the first networks are under deployment. In fact, there is already a need to prospect for new promising technologies, while identifying significant use cases for the next generation of wireless systems, which have been dubbed \ac{6G} communication systems. In the context of such future \ac{6G} wireless communication networks, this white paper focuses on the key aspects of the localization and sensing procedures by: 
\begin{enumerate}
    \item identifying potential enabling technologies and main features;
    \item assessing new opportunities of the environment-aware applications;
    \item recommending latest trends while posing key research questions.
\end{enumerate} 

Typically, wireless networks are praised for their communication features alone, while their inherent localization and sensing benefits are overlooked. In that regard, the \ac{5G} \ac{NR} access interface with its large bandwidth, very high carrier frequency, and massive antenna array offers great opportunities for accurate localization and sensing systems. Moreover, \ac{6G} systems will continue the movement towards even higher frequency operation, e.g., at the \ac{mmWave} as well as THz\footnote{While the THz terminology is appealing, it is incompatible with the common usage of \ac{mmWave}; at the same time, the term microwave is ambiguous. A good middle ground is to address the band of interest as ``\ac{muWave}''. Here, we use both terminologies interchangeably.} ranges, and much larger bandwidths. In fact, the THz frequency range offers great opportunities, not only for accurate localization but also for high definition imaging and frequency spectroscopy. In \cite{rappaport-ieee-access19} the authors provide an overview of the wireless communications and envisaged applications for \ac{6G} networks operating above \(100\)~GHz, and then highlight the potential of localization and sensing solutions enabled by \ac{mmWave} and THz frequencies. Along the same lines, possible directions for the cellular industry towards the future \ac{6G} systems are discussed in \cite{qi-ieee-cm19}. 

Location and sensing information in mobile communication systems has several applications, ranging from \ac{E911} emergency call localization to through-the-wall intruder detection, from personal navigation to personal radar, from robot and drone tracking to social networking. Location side-information can also be a service-enabler for communication network design, operations, and optimization. For instance, in \cite{taranto-ieee-spm14} and \cite{7984759} Taranto et al. and Koivisto et al. respectively overviewed location-aware communications for \ac{5G} networks across different protocol stack layers, and then highlighted promising trends, tradeoffs, and pitfalls. Additionally, the prospects for enhanced network synchronization are highlighted in \cite{7880669}. In \cite{rosado-whitepaper18}, the authors provide a general overview of localization methods for the \ac{5G} \ac{NR} and more specifically for the \ac{IoT} applications, i.e., they initially discuss key enabling technologies and features of \ac{5G} networks, and thereafter not only identify important practical implementation challenges, but also recommend potential development paths for localization-based services. In \cite{lohan-ieee-cm18}, Lohan et al. addressed how position side-information can be exploited to improve the operation of high-frequency \ac{IIoT} deployments. How these concepts, ideas, and services will change in the \ac{6G} era forms the core of this white paper.

The remainder of this contribution is organized as follows. In Section \ref{sec:techs}, we first identify key enabling technologies for the future mobile communications systems and highlight which desirable features are advantageous to localization and sensing. Thereafter, envisaged applications and opportunities are discussed in Section \ref{sec:apps}. Finally, in Section \ref{sec:questions}, we summarize our outlook for the future \ac{6G} localization and sensing development, while posing fundamental research questions.

\section{Enabling technologies for 6G environment-aware communication systems}
\label{sec:techs}

We can identify four emerging technological enablers for \ac{6G} communication networks, as well as localization and sensing systems. These are the use of new frequencies of the radio spectrum, the inclusion of intelligent surfaces, intelligent beam-space processing, in addition to \ac{AI} and \ac{ML} techniques. In this section, we cover these enablers in detail while discussing technological challenges and pointing out future opportunities. 

\begin{figure}
  \centering
  \includegraphics[width=.85\columnwidth]{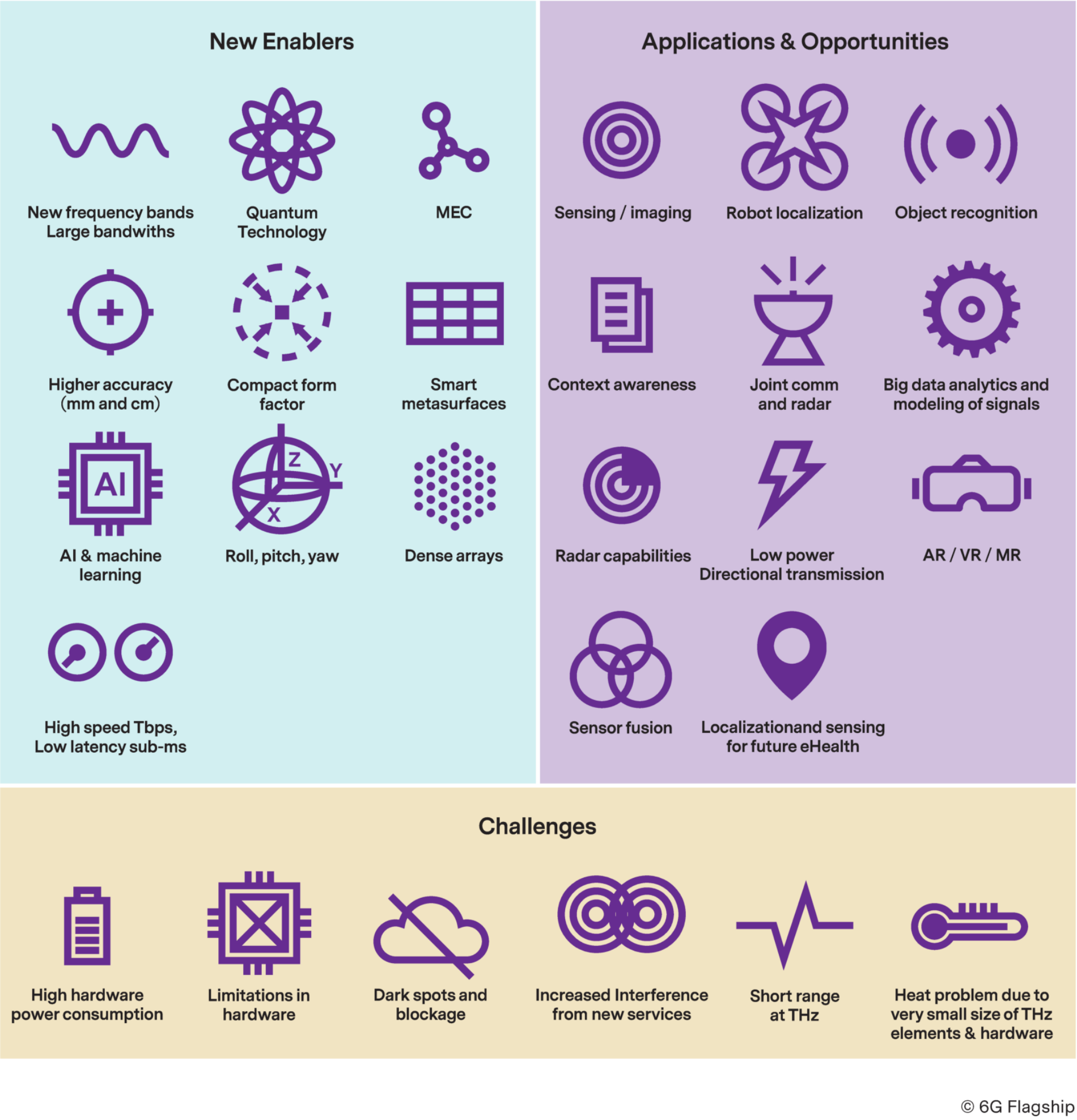}
  \caption{Chart relating enabling technologies, 6G new application opportunities and technological challenges.
  }
  \label{fig:6glandscape}
\end{figure}

Figure \ref{fig:6glandscape} summarizes key technological enablers and the resulting new opportunities for localization and sensing applications in \ac{6G} systems. In addition, the illustration identifies related challenges and open issues that still need to be tackled in order to take advantage of the new opportunities offered by much finer range, Doppler and angle resolutions and to realize the envisaged applications.

\subsection{\acs{RF} spectrum for future localization and sensing systems}
\label{sec:rf_spectrum}
 
In this section, we assess the prospect of using new \ac{RF} spectrum at high frequency ranges in \ac{6G} systems, while discussing the underlying radio propagation features and their impact (advantages and disadvantages) on future localization and sensing applications. We then briefly cover future chip technologies and developments in channel modeling. 
\subsubsection*{A leap in bandwidth and carrier frequency}
\begin{figure}[h]
  \centering
  \includegraphics[width=0.5\linewidth]{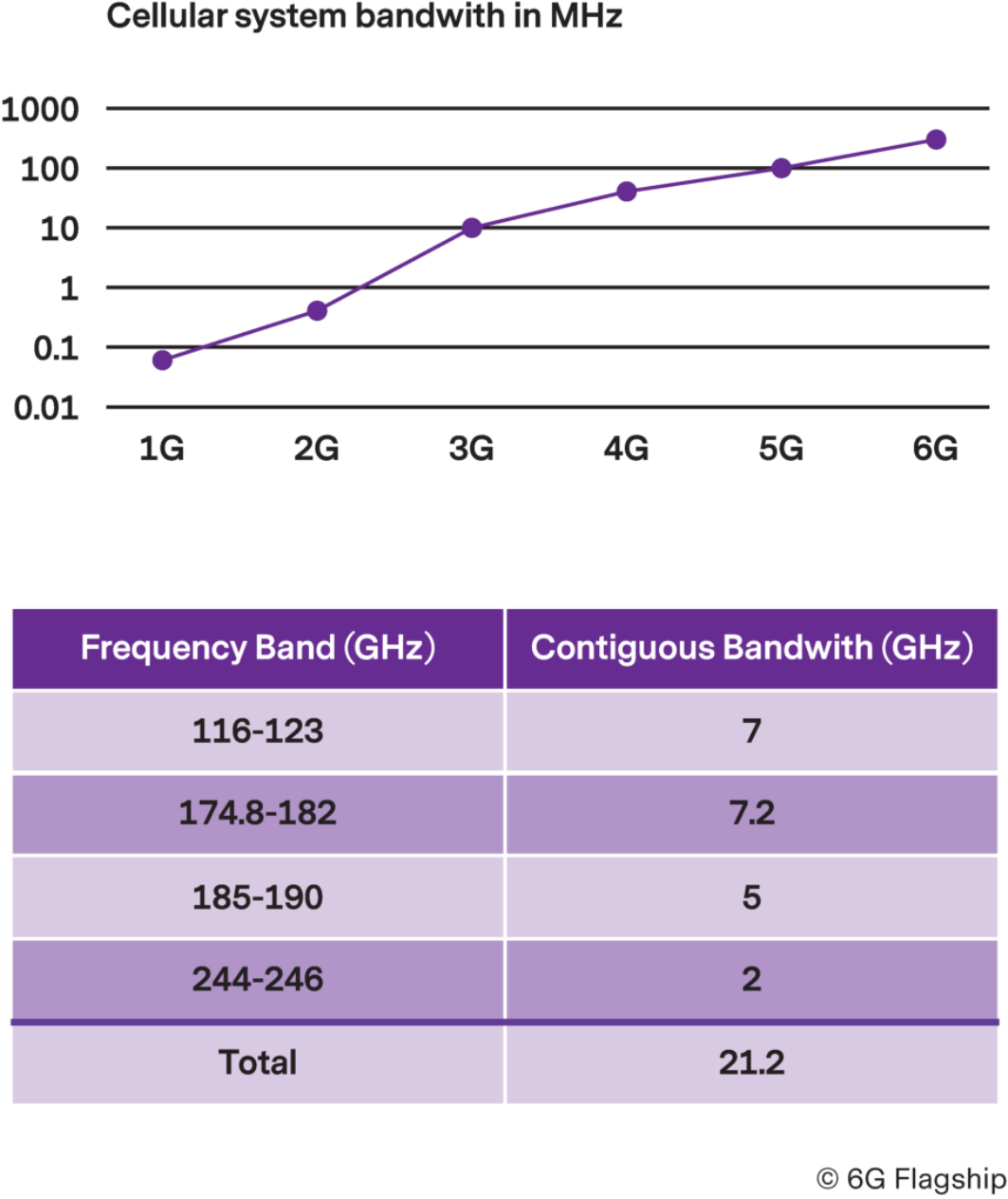}
  \caption{(top) Growth of channel bandwidth for cellular networks~\cite{qi-ieee-cm19}; (bottom) new unlicensed spectrum above \(100\)~GHz proposed by FCC~\cite{rappaport-ieee-access19}.
  }
  \label{fig:GenerationalBW_and_FCCproposal}
\end{figure}

As indicated in Fig.~\ref{fig:GenerationalBW_and_FCCproposal}, \ac{6G} radios will be likely to allocate services across channel bandwidths which are at least five times larger  compared to \ac{5G}, so as to accommodate ever-demanding data rates, increased reliability needs and demanding new services such as sensing and localization. Taking one step back, in up to \ac{4G}, only bands below \(10\)~GHz were occupied, whereas \ac{5G} \ac{NR} started implementing several so-called \ac{mmWave} bands.
Towards \ac{6G}, then, it is expected that the next-generation radios will thus aim to occupy higher bandwidths, now reaching above \(100\)~GHz. Furthermore,~\cite{rappaport-ieee-access19} (table reproduced in Fig.~\ref{fig:GenerationalBW_and_FCCproposal}) indicates the \ac{FCC} is on track to propose several new bands above \(100\)~GHz for unlicensed use. The key goal of doing this is to allow research and development activities towards these frequency bands. Regulatory activity is often indicative of a trend, and it is expected that these bands are likely candidates beyond \ac{5G} \ac{NR}.
Practical system and circuit evaluations today (for communications purpose) are already being tested at \(140\)~GHz \cite{qi-ieee-cm19,koziol-thz6g-spectrum19, rappaport-ieee-access19}. 
%

\begin{figure}[h]
  \centering
  \includegraphics[width=14cm]{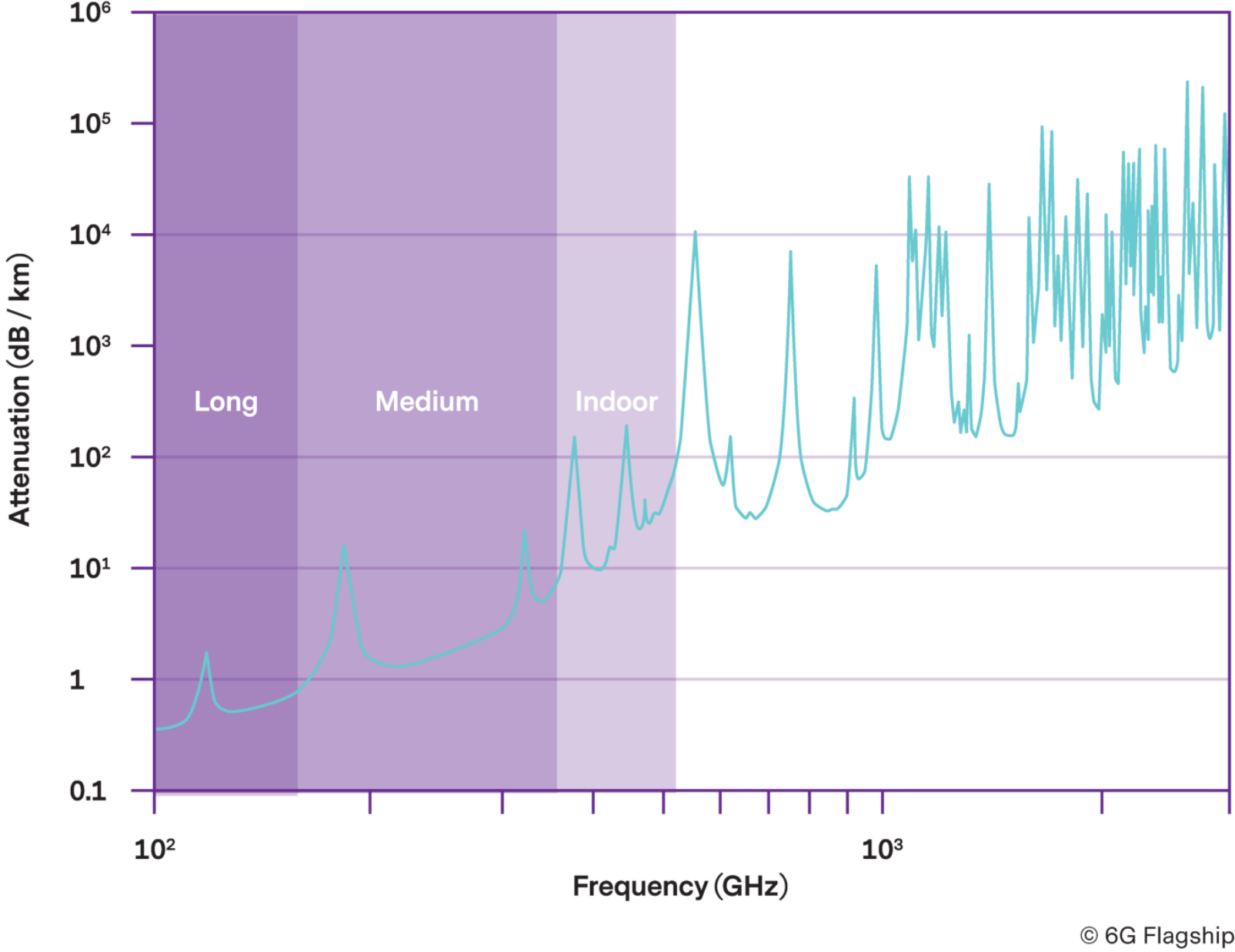}
  \caption{Illustration of the relation between frequency range and radio channel propagation effects (reflection, diffraction, scattering).}
  \label{fig:spectrumabsorption}
\end{figure}

Given the expected transition to higher frequencies and thus smaller wavelengths than used for \ac{mmWave} bands, the term \ac{muWave} is proposed. 
%
%
The increase in frequency has several important implications: \ac{mmWave} has a small beam width, typically referred to as a ``pencil beam'', while \ac{muWave} is expected to offer even more beam width reduction, leading to higher positioning accuracy, array gain in the link budget, and imaging capabilities. In addition, depending on the wavelength, there is a significant difference in absorption by the atmosphere when specific waves are used as illustrated in Fig.~\ref{fig:spectrumabsorption}. Hence, the \ac{EM} radiation can reach shorter distance at the higher frequency, e.g., \ac{mmWave} and \ac{muWave}.

From a localization and sensing perspective, the transition to THz frequencies has several important benefits. First of all, signals at these frequencies are unable to penetrate objects, leading to a more direct relation between the propagation paths and the propagation environment. Secondly, at higher frequencies, larger absolute bandwidths are available, leading to more resolvable multi-path in the delay domain with more specular components. Third, shorter wavelengths imply smaller antennas, so that small devices can be packed with tens or hundreds of antennas, which will be beneficial for angle estimation. In addition, the high-rate communication links offered by \ac{6G} will be able to be leveraged to quickly and reliably share map and location information between different sensing devices. This is beneficial for both active and passive sensing. To harness these benefits, chip technologies must be available that sufficiently support economies of scale. In addition, to support the development of new solutions and algorithms, suitable channel models that properly characterize the propagation of \ac{6G} waves over the hardware and the air are needed as well.

\subsubsection*{Future chip technologies}\label{sec:chiptech}

Having defined the broad initial range of relevant (new) spectrum for \ac{6G} to be as large as \(0.3\) - \(3\)~THz, while regulatory bodies have recently started to enable R\&D up to \(250\)~GHz, There is a clear need for further development of technology which will able to support the said frequency bands in a cost effective manner. One key aspect is the integration of the required technology. Currently, radio systems operating in the range of multiple \(100\)~GHz typically include antennas and signal processing equipment, for example, which is unreasonably large to integrate into typical \ac{UE}. 
As we start to migrate towards \ac{5G} systems in the world, we see that silicon products for \ac{UE} purposes have been in development for a few years while the network roll-out has started in a limited geographical scope only during \(2019\) and \(2020\). This is in no small part due to the complexity of the added air interface at \ac{mmWave} frequencies and increased usage of array technology to enable \ac{UE}-side beamforming. While the development of the front-end and modem chips has taken time, the process technology of choice is now available to enable efficient amplification, down-conversion and further analog processing, even at the large bandwidths. Planar bulk \ac{CMOS} technology at \(28\)nm has been able to achieve a so-called transition frequency at least one order of magnitude above the \(28\)~GHz and \(39\)~GHz operational bands.
Towards \ac{6G} operation, it is expected that a similar hurdle will have to be taken by chip technology. However, this time it is \emph{not} obvious that the process technology can handle the required frequencies. Even at ``only'' \(100\)~GHz, the same \(28\)~nm planar bulk \ac{CMOS} process would not be able to achieve the same efficiency e.g., to amplify signals. Hence, alternative technology options must be considered for cost-efficient solutions. %
Currently, multiple technologies are under study that have the potential to achieve good output power and efficiency at ever-higher frequencies. These include: \ac{GaAs}, \ac{GaN}, \ac{InP}, \ac{CMOS}, \ac{SiGe}, and \ac{FD-SOI} \ac{CMOS}. Today, the conclusion may be drawn that technically, operation up to \(300\)~GHz should be possible. However, above \(300\)~GHz, even \ac{InP} will no longer be able to provide efficient amplification and multipliers will have significantly lower output power to drive antennas \cite{GEMS_2020}. Hence, it is an open research question whether further technology development is required and if not, which technology is best suited for the \ac{6G} challenge.

\subsubsection*{Consistent channel models}

Electromagnetic wave propagation models are an important component for the proper design, operation, and optimization of wireless communication systems. In fact, when developing new algorithms or deploying new architecture, the simulation of radio channel dynamics and network operation becomes crucial for assessing the overall performance of mobile cellular systems. Propagation models are equally critical to ensure secure and reliable positioning, since it is impossible to design radar, sensing, ranging or direction estimation algorithms without sound hypotheses on how the electromagnetic waves propagate in the vicinity of the sensing device. Models are also critical to predict and benchmark the accuracy achieved by the various sensing and localization approaches. 
When compared to radio signals at lower frequencies, systems operating at high frequency are more susceptible to weather conditions, and do not propagate properly through materials. These properties are also relevant for sensing. For instance, for radar systems, the radar cross-section of different objects and the clutter properties of different environments must be characterized, and these must be correlated with the propagation models if we want to assess the performance of joint radar and communications systems. As we move into higher frequencies and into spectroscopic analysis, the absorption of electromagnetic waves through different gases and the reflection properties of different materials must be better understood and modeled.

In wireless systems operating at the microwave frequency range, the reflection and diffraction effects dominate the signal propagation and scattering is typically ignored. Differently, at \ac{mmWave} and THz frequencies the signal wavelength becomes comparable to, or even smaller than, suspended particles in the air (dust, snow and rain) or object surfaces irregularities and roughness. The path components resulting from the scattering may compensate for the radio channel degradation effects mainly as carrier frequencies increase above \(30\)~GHz. Moreover, typical drop-based statistical models \cite{3gpp-tr38901-18} fail to capture the targets' motion in highly directive antenna configurations. Thus, it is crucial to develop new channel models able to capture the \emph{consistency} of the propagation features.
\begin{itemize}
    \item Spatial consistency: A channel model is spatially consistent if its large- and small-scale parameters continuously vary with the target node position and depending on the reference geometry. In other words, nearby target nodes experience correlated large-scale features, while the small-scale parameters of each target (for example, angle or delay information) dynamically change with the position over the network deployment area. The authors in \cite{docomo5g} summarized relevant channel modeling considerations and noted that the proper characterization of the radio channel spatial consistency was the most challenging extension to be incorporate into traditional drop-based simulators. In \cite{3gpp-tr38901-18}, the 3GPP standardization body proposed a spatial consistency procedure for drop-based simulations that can be used with both cluster- and ray- random variables. However, it still misses implementation details related to capturing the correlation distances in highly directive narrow-band antenna arrays, as well as beam steering when considering high resolution scale.
    \item Frequency consistency: Consistency across frequency bands is also a desirable feature in channel models; low-frequency signals propagate over a wide area while enabling relatively coarse localization, while the higher frequency bands allow more accurate localization, albeit for a shorter range. This indicates that achieving both long-range and high-accuracy localization will require the joint processing of signals corresponding to widely spaced frequency bands, together with channel models faithfully linking the propagation characteristics across the whole spectrum. A specific model is required for channels involving large transmitting or receiving arrays or surfaces (massive \ac{MIMO}, distributed \ac{MIMO}). However, the channel model becomes non-stationary in space since certain parameters, e.g., path loss and \ac{AoA}, can not be assumed to be constant between the antennas \cite{DBLP:journals/corr/abs-1903-03085}.
    Consistency across frequency bands will also be critical for reliable localization in environments where the user might be simultaneously using several bands, e.g., if the network makes use of the \ac{CUPS} principle.
\end{itemize}

%

\subsubsection*{Challenges}

Finding the right band for the right frequency to optimize the use of potential new applications is a key challenge.
There is also the issue of the technology gap, i.e., how to make cost-effective and scalable systems towards \(300\)~GHz operating frequencies and beyond.
Future \ac{6G} systems will most probably introduce new use cases and various deployment scenarios which will require developing new (preferably general) dynamic radio channel models capable of capturing the time evolution of relevant metrics (such as channel impulse response) at a fine resolution. At a high frequency range, such channel models need to be consistent across frequencies and space, while capturing radio channel characteristics, material properties as well as features of the radio access interface. However, the channel coherence time will be significantly reduced and thus requires much faster and more frequent updates to properly capture radio changes. Moreover, measurement campaigns to either estimate parameters or validate models become difficult at such high frequencies due to, e.g., long measurement and processing periods, equipment costs and even presence/absence of foliage. In this configuration, ray tracing simulation is a well-known complement to field measurements, although developing accurate ray tracing simulators at a reasonable computational cost, given the necessary fine resolution time, also becomes a challenge.

\subsection{Intelligent reflective surfaces for enhanced mapping and localization}







\begin{figure}[h]
  \centering
  \includegraphics[height=0.7\linewidth]{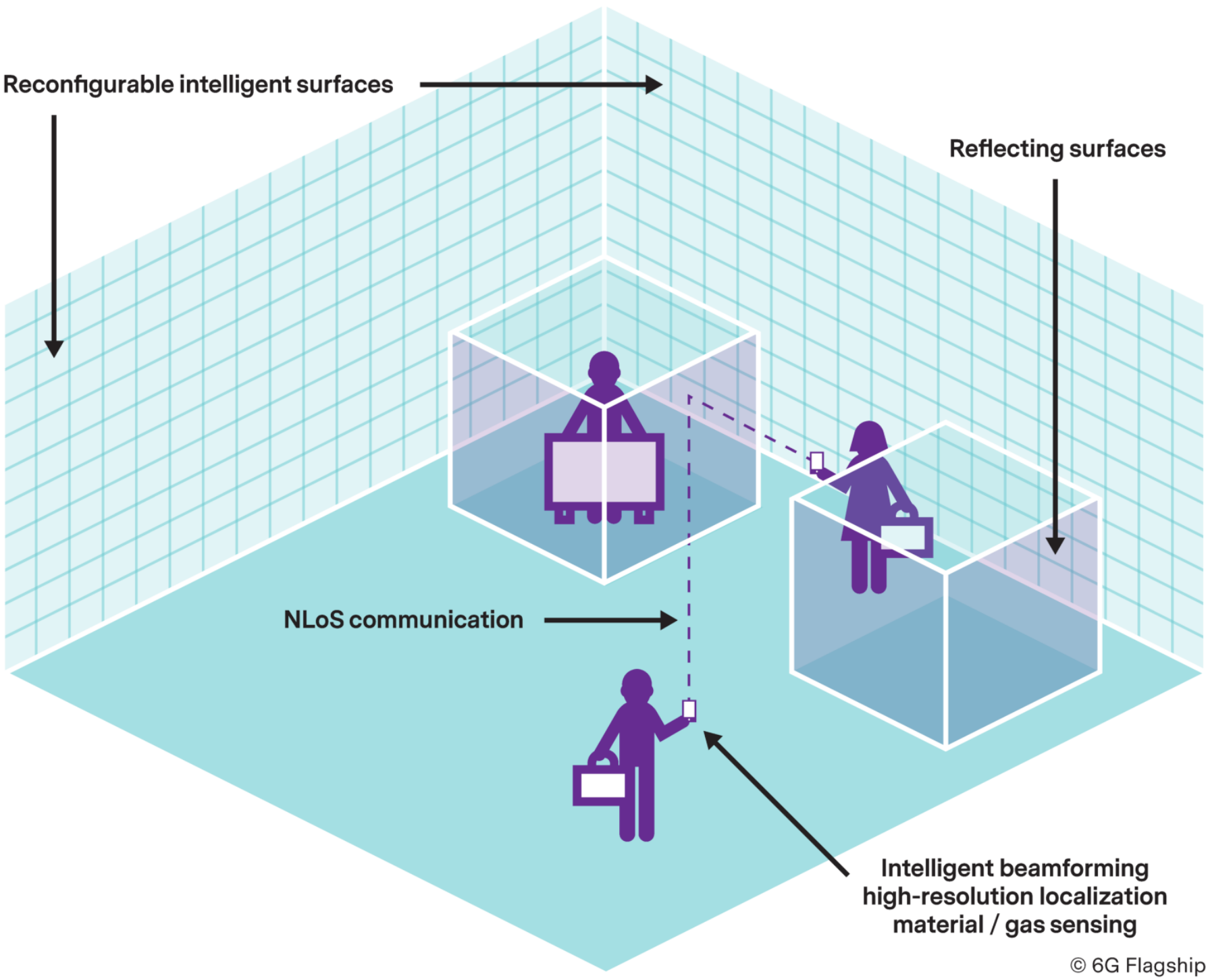}
  \caption{Illustration of IRSs in an indoor area, where the IRSs can facilitate the NLOS communication.}
  \label{fig:IRS}
\end{figure}

The radio propagation environment between transmitter and receiver peers has always been perceived as a random (non-controllable) component of wireless communication systems. Toward this end, \acp{IRS}\footnote{IRSs go under different names, including reconfigurable intelligent surfaces (RIS) and large intelligent surface (LIS).} have been recently introduced as a promising solution to control radio channel features such as scattering, reflection, and refraction.
%
In particular, \acp{IRS} allow network operators to shape and control the \ac{EM} response of the environment objects by dynamically adapting parameters such as the phase, amplitude, frequency, and polarization without requiring either complex decoding, encoding or radio frequency operations \cite{basar-ieeeaccess19}.
Succinctly, an \ac{IRS} is an \ac{EM} surface that is typically implemented through conventional reflective arrays, liquid crystals, or software-defined meta surfaces \cite{hu-arxiv19}. \ac{IRS}-assisted communications have the potential to enable low-complexity and energy-efficient communication paradigms. Even though \acp{IRS} with fixed \ac{EM} features have been previously employed in radar and satellite communications, it is only recently that they have found applications in mobile communication systems.
Figure \ref{fig:IRS} illustrates prospective application scenarios of \acp{IRS} in an indoor setting, where \acp{IRS} can extend the wireless communication range and facilitate \ac{NLOS} communications. An \ac{IRS} can reflect signals in precise directions by adjusting the phase shifter arrays, which can be controlled at access points/small-cell \acp{BS} via smart controller(s). Hence, \acp{IRS} are effectively large-scale arrays of phase shifters that do not have their own radio resources. Consequently, an \ac{IRS} cannot send pilot symbols to help a small-cell \acp{BS} to estimate the channel response between small-cell \acp{BS}-\acp{IRS}. 

The importance of \acp{IRS} is particularly emphasized at high frequencies in the context of \ac{mmWave} / THz communications \cite{sarieddeen2020overview}. With larger path and penetration losses and lower scattering at higher frequencies, the number of naturally occurring propagation paths is low. Furthermore, regular large coherently-operating antenna arrays are difficult to implement at high frequencies (as the size of antenna elements gets smaller), and the relaying technology is also not yet mature. \ac{IRS} deployments can thus solve these issues, where the addition of controlled scattering can extend the communication range and enhance system performance. Note that electronically-large (large compared to the operating wavelength) \acp{IRS} at high frequencies can be realized with very small footprints, which further facilitates their deployment.

The aforementioned intelligent operations of \acp{IRS}, alongside the inherent \ac{EM} features, especially in the \ac{mmWave} / THz -band, can further enhance the performance of localization and sensing. For instance, an \acp{IRS} could enable tracking/surveillance applications in \ac{NLOS} communications and autonomous localization. If the intelligent surfaces are large, the near-field effects provide an opportunity for exploiting the wavefront curvature, which can improve location accuracy and possibly remove the need for explicit synchronization between reference stations \cite{hu2018beyond}. On the other hand, accurate sensing and localization (especially at the THz band) can, in turn, enhance the efficiency of \ac{NLOS} signals and reflective surfaces, whether intelligent or not (as illustrated in Fig. \ref{fig:IRS}). In particular, intelligent surfaces operating at the THz band can be very small, and hence accurate localization techniques would be required to track their relative location in real-time. Furthermore, the material type of the reflecting surface affects the efficiency of \ac{NLOS} signals. For example, regular (non-intelligent) smooth surfaces introduce a significant specular reflection component in the THz band. Passive, reflection-based THz sensing can determine the material type of surfaces, which results in intelligent beamforming decisions. This is in line with the concept of leveraging the knowledge of the environment for enhanced communications, sensing, imaging, and localization applications \cite{sarieddeen2019next}.


\subsubsection*{Challenges}
Several technological challenges and open issues still need to be addressed to make \acp{IRS} a viable alternative in future \ac{6G} systems. To carry out research and push the further development of \acp{IRS} forward, appropriate models that describe the properties of the constituent materials and the radio propagation features of the incident signal waves are still lacking. Feasible implementations (accounting for hardware and software aspects) of such reconfigurable meta-surfaces are also essential to ascertain their potential gains in wireless communication systems. \acp{IRS} also need to collect the \ac{CSI} between communicating peers to adjust the radio characteristics of the impinging signals properly. Therefore, it is necessary to devise novel energy-efficient methods to estimate the channel properties of the wireless links. When considering the integration of such \acp{IRS} into the infrastructure of the wireless systems, it will also be necessary to develop strategies to identify adequate locations for their deployment in the network coverage area. Furthermore, novel signal processing techniques that optimize the performance of \ac{IRS}-assisted joint communications, sensing, and localization are required. This is especially challenging at high frequencies, where the overall channels tend to be of low-rank, and hence carry less information.

\subsection{Beamspace processing for accurate positioning}

\begin{figure}[h]
  \centering
  \includegraphics[width=0.7\textwidth]{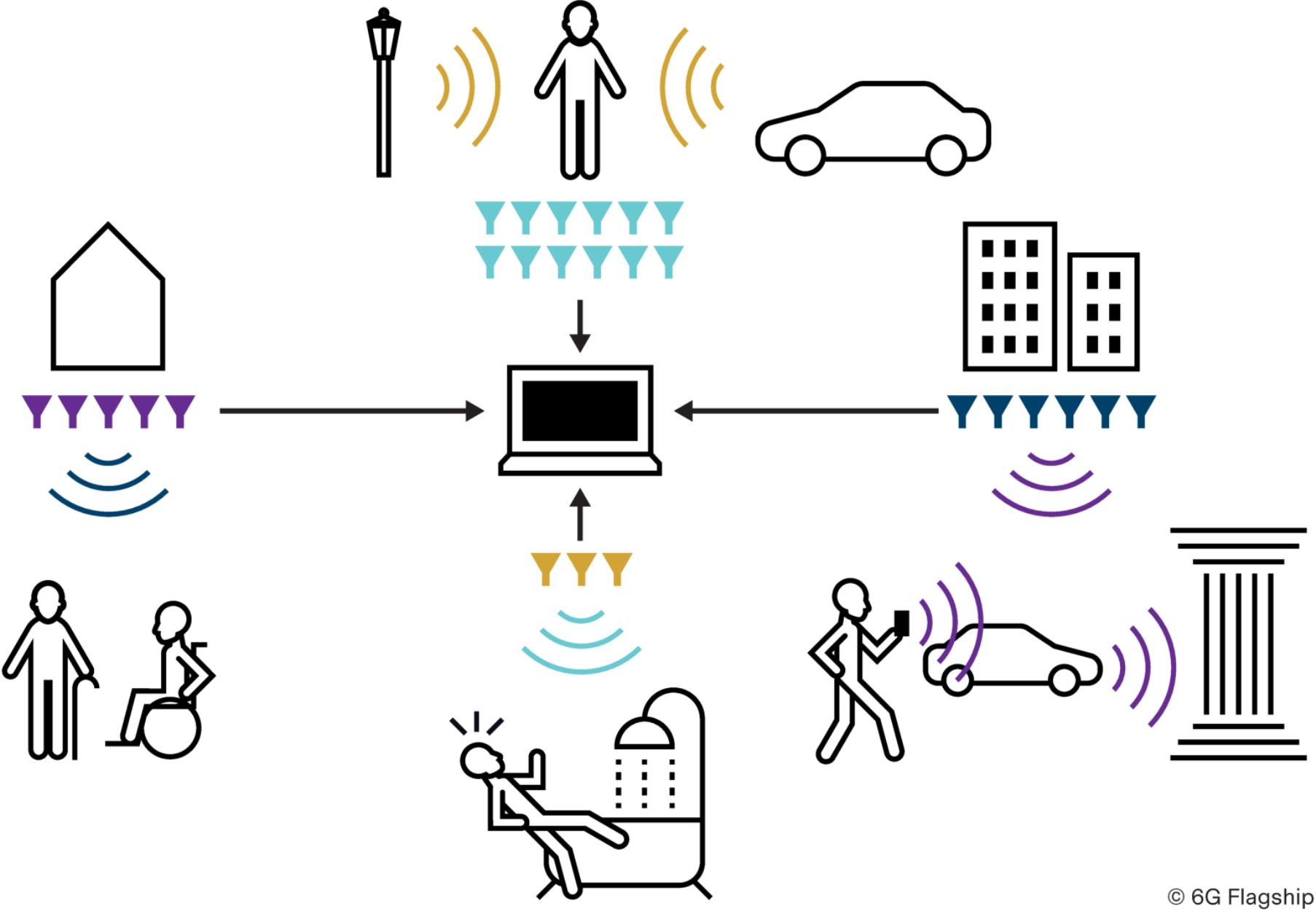}
  \caption{Illustration of the beamspace domain/processing for accurate positioning. This figure depicts the collaborative multi-antenna systems deployed in target space with distributed topology for managing hybrid beamspace and localizing passive and active targets; the link ensured between the multi-antenna systems and the active target could be the LOS and NLOS paths and could be blocked by moving background objects.}
  \label{fig:beamspaces}
\end{figure}

Beamspace is one of the promising enablers for localization and sensing in \ac{6G}. 
The beamforming at \ac{mmWave} and \ac{muWave} is essentially the transmission of coherent signals thus forming a concentrated field in a certain direction to increase the signal-to-noise ratio or throughput as a beamforming gain. Enhanced beamforming capacity in \ac{3D} space is advantageous to overcome the high path loss in \ac{mmWave} and \ac{muWave} bands and to mitigate the interference from different directions by forming very narrow beams. A beamspace channel response collected by a monostatic or multistatic transmitter and receiver contains spatial information of not only the link ends but also the interacting objects/humans in between, which can be processed for localization and sensing purpose.
Beamforming can be categorized into analog, digital, and hybrid beamforming \cite{ahmed2018survey}, where the former supports a single stream and phase shifters allow directional transmission (depending on the aperture), while the second option allows multiple streams in different directions. 
Beamspace processing relies on advances in channel estimation, including angular and delay domain profiles, needed for localization and sensing algorithms. The channel estimation is particularly critical in unfavorable \ac{NLOS} and high mobility scenarios.
As illustrated in Fig.~\ref{fig:beamspaces}, in the case of localizing and tracking active mobile users in a changing environment, the beams are managed dynamically as the channel estimation of the \acf{AoA} or \ac{AoD}. 
There exist a plethora of methods for \ac{AoA} or \ac{AoD} estimation, with varying degrees of complexity and performance. 
In terms of the localization of active mobile users who can be connected with the base station by a \ac{LOS} path, or strong \ac{NLOS} path such as specular reflection, accurate and fast \ac{AoA} or \ac{AoD} estimation of the major multipath is sufficient.
In the uplink direction, \ac{AoA} estimates from \ac{LOS} links allow to infer the location of the active user directly, while \ac{NLOS} links, resulting from multipaths of the signal transmitted by an active user boucing off scatters and then arriving at the base station, allow to estimate the location of a scatterer (interacting object).
In the case of the latter, i.e., the \ac{NLOS} scenarios, in order to trace down the location of the user, both \ac{AoA} and \ac{AoD} estimations are needed.

In addition to beamspace processing for the localization of active users in \ac{LOS} and \ac{NLOS} scenarios, it is even more challenging if the target does not carry or wear any device, i.e., device-free localization \cite{Devicefree1} or sensing.
In this case, there is a need for additional environment indications where the spatial characterization of the target(s) can be distinguished from the background objects.
This is especially challenging when multiple targets are present \cite{MultihumanLoc2} and the identification of the targets \cite{IdentificationHumanss} is needed as well.
With a pencil-like beam operating at \ac{mmWave} and \ac{muWave} frequency bands, the spatial resolution is very high. At the ultra-wide bandwidth, the delay resolution is also very high. The combined angular-delay profiles of beamspace channels serve to localize and sense passive targets.
The beamspace channels can be collected continuously and in this case the instantaneous beamspace can be compared with a reference (collected off-line to represent the static environment) to sense a passive target, and compared with a previous sample (on-line or real-time) to track a moving target. 
The identification of targets needs the help of learning algorithms to distinguish the angular-delay profile variations of different target (present in different size, dielectric properties, etc.).
Dedicated beamforming signals and the deployment of strategic monostatic/multistatic \ac{MAS} could support maximum accuracy for the localization and sensing of scattering targets.

\subsubsection*{Challenges}
An important challenge in beamspace processing is blockage, e.g., due to high mobility background objects cause deep fading and thus influence the localization accuracy. 
In case of deep fading, the localization and sensing function of a current serving \ac{MAS} may be temporally switched to another \ac{MAS}.
In order to coordinate the available \ac{MAS}s deployed in a target area for real-time localization and tracking, the prevention of blockage from moving background objects which can cause dramatic degradation of beamspace signal quality is necessary.
This could be achieved through image-based mobility-behavior prediction, geometrical domain environmental identification and radio channel simulation, ideally in a runtime in the order of milliseconds.
In high mobility scenarios, the observation of moving background objects which could potentially block the beamspace could be implemented, e.g., using a depth camera; the learning and prediction of its moving trajectory can be conducted in real-time; the potential influence on the beamspace channels could also be predicted by ray tracing simulations or a hybrid method of ray tracing and a propagation graph \cite{HybridRTPG2}. 
Additional practical challenges include the following.
First, with beamforming at higher frequencies more than \(40\)~GHz (post-\ac{5G} frequencies), there are risks of increased phase noise and non-linearity of systems, which could affect the signal quality and result in channel estimation errors and hence influence the functionality. 
Second, when the target is in the near-field of a multi-antenna station, It will be necessary to devise direction estimation algorithms for beamspace channels for a robust estimation. 
Third, how the beamspace separation influences the accuracy of identifying multiple targets that are closely located together is yet to be investigated thoroughly.
Fourth, the \ac{AI}-enabled precoding for digital beamforming combined with a dynamic multipath simulator with environment information could add constructively for even sounder system attributes. Nevertheless, the trade-off between the algorithm complexity, the hardware capability and the time consumption needs to be balanced. 
A combination with learning algorithms for target recognition is promising; For this purpose, comprehensive performance evaluation matrices, current and foreseeing localization accuracy, and the Cram\'er-Rao bound on estimation accuracy need to be defined with the specifications of the system configuration.


\subsection{Machine learning for intelligent localization and sensing}\label{sec:ai_ml}

\begin{figure}[h]
  \centering
  \includegraphics[width=.85\columnwidth]{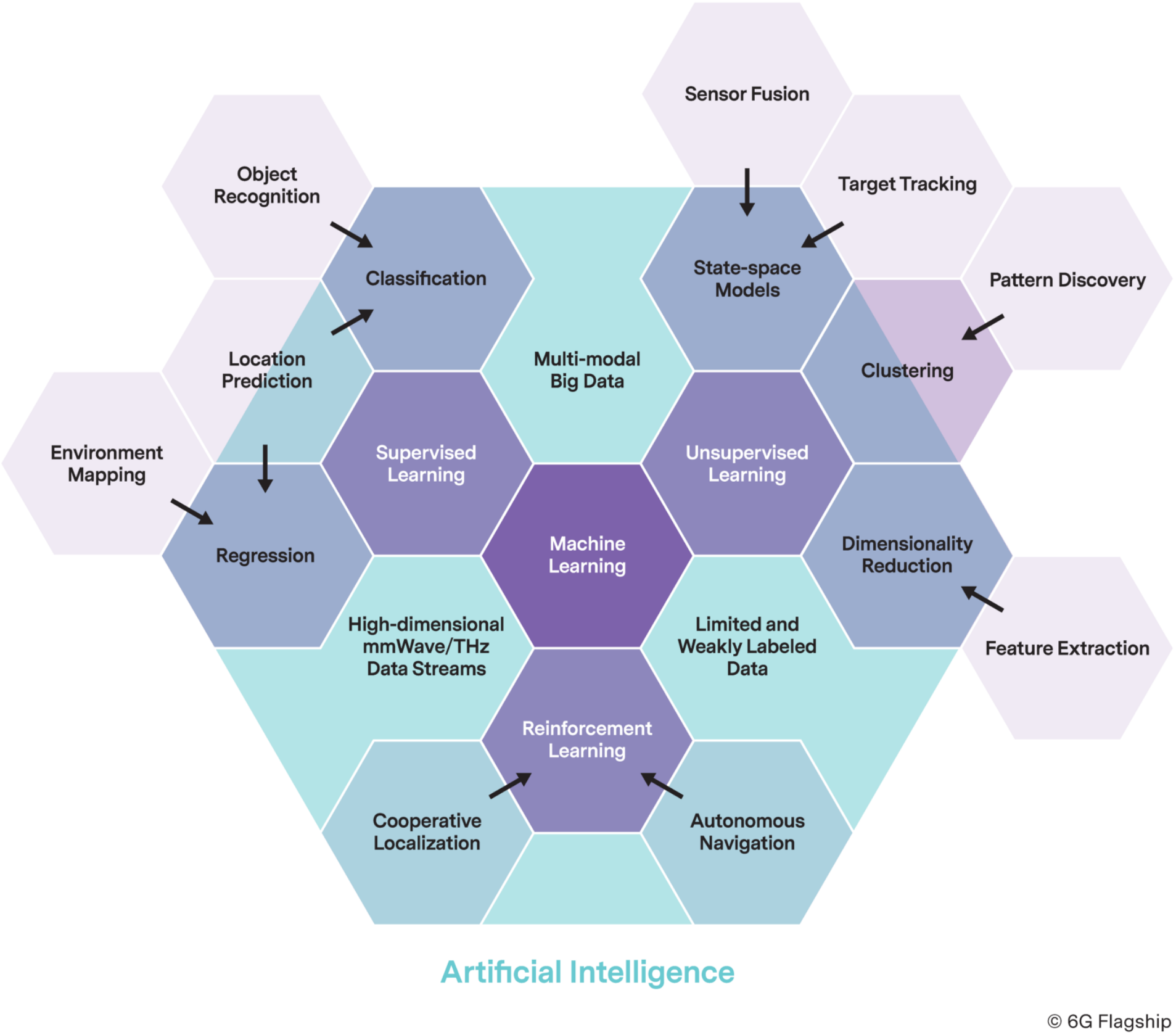}
  \caption{Landscape of the \ac{AI}- and \ac{ML}-based approaches for the localization and sensing solutions.}
  \label{fig:ml}
\end{figure}

\ac{AI} techniques are becoming ever more important moving towards the data-rich \ac{6G} era. Studying how to build intelligent systems and agents which are able to achieve rational goals based on logical and probabilistic reasoning, planning and optimal decision making, possibly in uncertain environments is a broad field. Modern \ac{AI} systems are typically based on \ac{ML} \cite{murphy2013}, which provides data-driven multidisciplinary approaches to learn models beyond explicitly programmed rules. 
\ac{6G} systems and beyond will rely on such data-driven algorithms, providing new opportunities not only for wireless communication but also for advanced localization and sensing techniques operating at the \ac{mmWave} and \ac{muWave} frequency ranges.

\subsubsection*{Localization}
\ac{ML} methods in localization mainly focused on fingerprinting and the usage of regression and classification methods \cite{hsieh, Zafari2019, Bekkali2011}. 
In data-rich and complex localization applications, especially for the \ac{GNSS} poor indoor and urban outdoor channel conditions, we expect a more widespread use of \ac{ML}, since traditional mathematical models and signal processing techniques are not alone able to solve challenging problems where we have a large number of multi-modal, indirect and noisy observations, and the physical properties of non-linear signal characteristics of the system are possibly unknown or difficult to model. Instead, we can utilize \ac{AI} methodologies to model how 
 the system behaves, including sensor noise and different uncertainties of the system. Furthermore, in many cases, to realize high-level sensing and localization from possibly high-dimensional low-level raw measurements, e.g., \ac{CSI} in massive \ac{MIMO} systems, predictive models and pattern recognition based on \ac{ML} techniques are essential. 
Hence, the use of statistical \ac{AI} methods to simultaneously model complex radio signal characteristics and fuse many complementary yet noisy sensors will become more important in future localization and mapping systems. These will be likely to be supported by hybrid models combining traditional physics-based models of signal propagation with data-driven learning approaches and sequential Bayesian state-space models \cite{murphy2013}. 
Furthermore, as the predictive function of mapping low-level measurements to high-level target concepts becomes extremely complicated, it will be impossible to build the mathematical model by hand. Machine learning provides an alternative framework which is able to learn from data by optimizing or inferring unknown free parameters (or latent variables) of the model to build more flexible and accurate approaches to \ac{6G} localization and sensing, based on state-of-the-art deep learning and probabilistic methods \cite{murphy2013}, for instance.

\subsubsection*{Sensing}
\acs{RF}-based sensing at high carrier frequencies will provide more accurate techniques to measure the environment, detect and recognize objects, and the wider spectral range will provide opportunities to sense and identify new kinds of targets and variables which are not detectable in currently used frequency bands \cite{rappaport-ieee-access19}. As data becomes more highly-dimensional and complex, \ac{AI}- and \ac{ML}-aided, together with these novel sensing capabilities, will provide opportunities not seen before. New kinds of high-level information and patterns hidden in the raw data will be extracted and many weak and noisy signals will be integrated temporarily and spatially to realize novel sensing approaches. On the one hand, extracted patterns and (latent) variables can be seen as a virtual sensor, providing an intermediate step to performing higher-level reasoning, for example, to realize more precise target localization. On the other hand, \ac{ML} algorithms can be trained directly to predict the characteristics of a dynamic environment. For example, this may be applied to detecting and classifying objects in variable conditions \cite{Tapia2019} or identifying users \cite{Zhao2019} and recognizing user behavior and contexts to enable future wireless communication networks as well as novel services and applications using passive sensing (cf. Section \ref{sec:passive_sensing}).

Target applications of \ac{ML} aided localization and sensing can thus vary from low-level feature extraction and pattern discovery to object detection and recognition, location tracking and prediction, environmental mapping, cooperative localization, channel charting (see Section \ref{sec:channelcharting}), and autonomous navigation and planning, for instance. Fig.~\ref{fig:ml} illustrates the \ac{AI} and \ac{ML} landscape of general supervised, unsupervised, and reinforcement learning concepts and their relations. Furthermore, a set of methods and models in each concept relevant to the localization and sensing as enabling technologies are shown as well as typical challenging and opportunistic data types and applications arising from the area of future \ac{6G} systems.

\subsubsection*{Challenges}
The present-day \ac{AI} and machine learning toolbox already provide a variety of techniques to extend traditional localization and sensing, for example, based on uncertainty-aware probabilistic learning and reasoning methods and deep neural networks \cite{murphy2013}. However, many of these well-established techniques have their limitations because they are data-hungry, requiring a large amount of labeled training data and computing power. To be able to learn from limited, arbitrary structured, and noisy data, novel hierarchical models and advanced inference techniques need to be developed and combined in clever ways. To overcome the cost of collecting labeled training data, several recent approaches could be further developed. Semi-supervised learning combines a small number of labeled data points with a large number of inexpensive unlabeled data points to refine a supervised solution. Furthermore, the characteristics of localization and sensing in next-generation communication systems will be more autonomous, often non-stationary and time-evolving, requiring online adaptive \ac{ML} techniques. For instance, federated learning-based crowd-sourcing \cite{ciftler2020} and reinforcement learning-based cooperative localization \cite{Peng2019} could help to overcome the challenges of limited data and adaptive environments. To be able to apply \ac{AI} and \ac{ML} techniques successfully as an enabler for the next-generation highly dynamic large-scale localization and sensing systems, solving these challenges will become essential.



\section{Localization and sensing opportunities for future 6G systems}
\label{sec:apps}
In this section, we discuss new localization and sensing opportunities, which are enabled by the aforementioned key technologies and are tailored for the upcoming \ac{6G} wireless communication systems. In this context, it is worth emphasizing that localization, sensing and communication must all coexist, sharing the same time-frequency-spatial resources in the envisioned \ac{6G} systems. There are different mechanisms to enable such sharing, including: coexistence, cooperation, and co-design \cite{chiriyath2017radar}. In this sort of sharing, mutual interference can be a challenge if not properly addressed. 
While any form or sharing can be interpreted as reducing the performance of at least one functionality, the interplay is not necessarily a zero-sum game: sensing and location information can guide communication (e.g., for beamforming or for hand-overs), while communication can support localization and sensing by sharing map information between devices. Sharing is not limited to the resources but can also exist at a waveform or hardware level. In terms of waveforms, communication waveforms may have undesirable properties from a localization or sensing point of view and vice versa. The design of joint parameterized waveforms will alleviate this problem. In terms of hardware, the same resources will be used for all three functionalities of localization, sensing, and communication.



User requirements and \ac{KPI} defined for \ac{5G} positioning services are equally important and have been under active discussion in the 3GPP releases \(16\) and \(17\).
However, it is worth noticing that there is a significant gap between the most stringent requirements of the use cases, as described in \cite{3gpp-tr22872-18,3gpp-ts22261-20}, and what has been effectively achieved in the current 3GPP standards, in particular, in Release \(16\) \cite{3gpp-tr38855-19,keating2019overview}. This gap between the requirements of the envisaged use cases and the performance targets in Release \(17\) may become a clear objective to be attained by the future \ac{6G} systems. Furthermore, \ac{6G} should include \acp{KPI} related to the new functionalities that it will include. In particular, the \ac{UE} orientation and the map or features of the environment should be captured in the corresponding accuracy metrics. \acp{KPI} characterizing the power consumption per position fix should also be considered since this is an essential aspect for positioning in the \ac{IoT} context. 
Besides flexibility and availability, \ac{6G} systems/protocols/signals should be designed so that the user equipment performing or supporting localization using \ac{6G} should also be able to continuously monitor integrity (to detect, isolate, and exclude signals from faulty transmitters or those heavily affected by multipath). In this context, some \acp{KPI} similar to legacy definitions of integrity monitoring such as time-to-alarm, false alarm rate, the maximum number of false transmitters that can be detected, etc. should be considered.

\begin{figure}[h]
  \centering
  \includegraphics[width=0.7\linewidth]{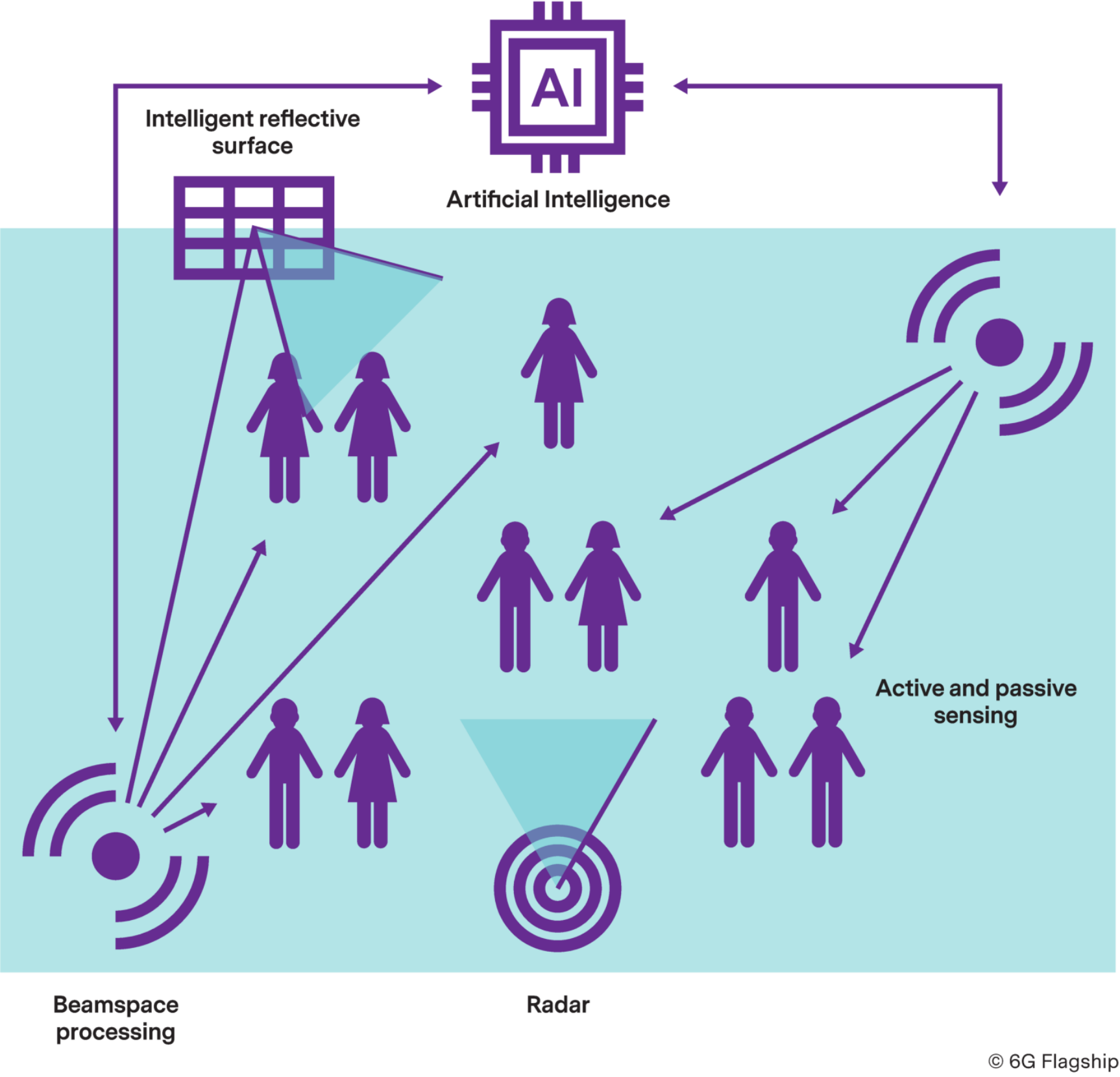}
  \caption{Illustration of the envisaged 6G opportunities and applications.}
  \label{fig:enablers}
\end{figure}

Figure \ref{fig:enablers} illustrates future deployment scenarios wherein the aforementioned technological enablers are employed to create exciting new opportunities for localization and sensing applications. In the following sections, we go into details and discuss various such application and service opportunities which will become a reality in the convergent \ac{6G} communication, localization and sensing systems.

\subsection{{THz} imaging} \label{sec:imaging}

The higher frequency bands, encompassing the mmWave and THz bands, offer unique opportunities for sensing because they allow very fine resolution in all physical dimensions: range, angle and Doppler. Both active and passive sensing and imaging are possible, where a passive sensor exploits the emissivity or the natural reflection of surfaces and uses an array of imager pixels to capture the image, in a non-coherent way, just like a conventional camera but in a much lower part of the electromagnetic spectrum. In contrast, an active sensor transmits a carefully designed sounding waveform and processes the echoes coherently to extract range, Doppler and angle information at a high accuracy and resolution. Both will now be discussed in more detail, followed by an introduction to an emerging application area: biomedical sensing and imaging.

\subsubsection*{Passive imaging}
The THz imaging state of the art reports two main competing categories of THz \ac{2D}-array image sensors. On the one hand, there is the above-IC bolometer\footnote{A bolometer is a device that measures the power of incident \ac{EM} radiation via the heating of a material with a temperature-dependent electrical resistance.} based THz image sensors based on a classical \ac{IR} sensor, which offers high sensitivity and currently has a good maturity \cite{simoens2014terahertz}. However, using two circuits (the sensor layer and the \ac{CMOS} circuit for the data extraction) and the necessary above-IC technology for assembling the layers make bolometer based image sensors expensive. On the other hand, monolithic \ac{CMOS}-based THz imagers have recently emerged as a low-cost competitor \cite{boukhayma2016low, al20121}. For these images, two kinds of architecture exist. First, the THz detection can be carried out by heterodyne demodulation. However, due to the pixel architecture and its high power consumption, such an imaging pixel cannot be used in a \ac{2D}-array sensor, and its use is limited to raster scanning techniques. The other possibility is to use incoherent (or direct) detection using a simple MOS transistor, resulting in a pixel with low complexity and power consumption. The THz antenna and its MOS detector provide a low-frequency output signal, proportional to the THz wave, which is amplified prior to readout. This property allows using a quasi-classic image sensor readout scheme which is fully compatible with \ac{2D}-array sensors. Even with their poor current sensitivity (\(1000\) times less than bolometer-based sensor), this MOS-based THz image sensors have proven to be a viable cost-effective alternative to bolometer-based imagers.

\subsubsection*{Active imaging}
We envision two distinct application areas for active imaging: active radar and material sensing. 
%
    Active radar imaging makes it possible to add the range and even Doppler dimensions to the image (\ac{3D} or \ac{4D} imaging). On the lower edge of the spectrum, in the \ac{mmWave} and low THz bands, radar imaging is evolving fast to satisfy the requirements of \ac{ADAS} and autonomous driving. The trend there is to resort to \ac{MIMO} techniques whereby a virtual antenna array is created with a size equal to the product of the number of transmitting and receiving antennas. \(79\)~GHz radar imaging with a wide field-of-view, resolutions of 
    \(1\deg\) by \(1\deg\) and a cm-scale range resolution is experimentally feasible today, and radars with a wide field-of-view and \ac{LIDAR}-like resolutions are an active field of applied research. Using higher carrier frequencies such as \(140\) or \(300\)~GHz is a longer-term trend, resulting in a smaller form factor or better angular resolution as well as better range resolution, thanks to the wider bandwidths. Some experimental systems already show the potential of \ac{CMOS} in the low THz regime (\(140\)~GHz) \cite{Visweswaran2019}.
    Active radar imaging has interesting applications beyond automotive radar such as body scanning for security, smart shopping and gaming. For shorter range applications, antennas can even be integrated on-chip \cite{Visweswaran2019} in bulk \ac{CMOS} for ultra-low form factor gesture recognition, vital sign monitoring and person detection and counting. It is expected that SiGe or III-V compounds will complement \ac{CMOS} for the \ac{RF} part when the carrier frequency is higher than about \(200\)~GHz \cite{Statnikov2013}, \cite{Ahmed2018}.
    
In addition to active radar sensing, 
    the vastly wider channel bandwidths, narrow beams and compact antenna arrays at the THz band also enable material sensing.  
    Fundamentally, it is the THz-specific spectral fingerprints that many biological and chemical materials possess that brings a great deal of potential to THz wireless sensing. For instance, THz rays can be used to study water dynamics by analyzing molecular coupling with hydrogen-bonded networks, as well as to monitor gaseous compositions via rotational spectroscopy. THz \ac{TDS} is typically used in such sensing applications, which consists of probing a material/medium with short pulses, the frequency response of which covers the entire frequency band, and recovering the absorption or transmission coefficients at the receiver. Following signal acquisition, a variety of signal processing and ML techniques can be used to pre-process the received signals, extract characteristic features, and classify the observations. %
    As an alternative to THz-\ac{TDS}, and with the advent of THz technology, carrier-based sensing setups can be used, assisted by ultra-massive \ac{MIMO} configurations. For example, each subset of antennas that is fed by a single \ac{RF} chain can generate a narrow and directed beam at a specific target frequency. With such fine-tuning capabilities, only a select few carriers can investigate the components of a medium by selecting these carriers to be close to the resonant frequencies of target molecules, for example. In such a carrier-based setup, THz sensing can be piggybacked onto THz communications \cite{sarieddeen2019next}. However, many challenges need to be addressed, from a signal processing perspective to enable efficient joint THz sensing and communications.

\subsubsection*{Biomedical applications}
Thanks to advances in semiconductor technology, packaging and signal processing, THz imaging is gaining interest in a number of application areas \cite{Mostajeran2017}. Healthcare has the potential to be transformed by digital health technologies, including highly compact and wearable biosensors, which are garnering substantial interest due to their potential to provide continuous, real-time physiological information via dynamic, non-invasive, contactless measurements. \ac{MEMS} sensors (e.g., inertial measurement units, pressure and temperature sensors) are now widespread in consumer devices (e.g. smart phones and smart watches), and both these and heart rate sensors are becoming more widely used in medical applications. However, there is a need for more radical mobile sensor technology to provide passive, continuous, home-based monitoring of biochemical markers in biofluids, such as sweat, tears, saliva, peripheral blood and interstitial fluid. THz imaging and spectroscopy has the potential to provide such a data source for future digital health technologies. For example, the measurement of chemicals in sweat has great potential to impact diverse medical applications, including wound healing, metabolic activity, inflammation and pathogens, and proteomics.
Current commercial and research THz systems which are suitable for biomedical applications (such as Teraview \cite{teraview} and Terasense \cite{terasense}) have a high cost and use bulky optical components. THz integrated \ac{CMOS} design trends including the source and essential capability of beam steering in a module or by extension a \ac{SIP}, will provide highly compact sources that leverages existing advanced electronic manufacturing processes to deliver a product at a small fraction of current production price points. The compact size of the sources would also reduce other product costs including distribution and installation.

\subsubsection*{Challenges}
THz sensing and imaging target small defects in materials thanks to the small wavelength of THz frequencies for industry market;  clinical, bio and agro-food analysis thanks to \(\text{O}_2\) or \(\text{H}_2 \text{O}\) peaks of absorption in the spectrum. The technology is also useful for security imaging thanks to the contrast between water-based tissues and other materials. This poses specific challenges for the transmitters and the receivers. 

Challenges for transmitters: Specific \(\text{H}_2 \text{O}\) absorption peaks are at THz frequencies, except the one at \(180\)~GHz. To create contrast, the radio source must provide radiated power and a focused beam. The available power gain depends on the Fmax value (which yields the frequency when the transistor power gain is \(0\)~dB) of the technology, and the available power output depends on the breakdown voltage of the transistors; the trade-off between these characteristics must be solved by silicon technologies to reduce the size and cost of applicative solutions. The focused beam can be obtained by antenna arrays, but the integration of these objects is a challenge. In addition, the orientation of the beam requires phase array functions, which at THz frequencies, requires either active solutions, and we come back to the Fmax-BV (Breakdown voltage represents the maximum DC voltage before the breakdown of the transistor) trade-off, or passive solutions using a new material phase shifter, with through path loss and integration challenges.

Challenges for receivers: The sensitivity of passive imagers, based on the illumination of MOS passive transistors, is not optimized. The technological challenge in the next years will be to optimize the sensitivity of the transistors to THz sources, and to reduce the size of these transistors, in order to increase the image definition and contrast. The sensitivity of coherent receivers (active), mainly for radar applications, is also a big challenge. This is defined by the \ac{NFmin} of the technology, and the \ac{LO} phase noise at THz frequencies. The process technology \ac{NFmin} must be reduced at very high frequencies, while the design challenge will focus on the \ac{LO} generation, even if the technology selection determines the \(1/f\) cut frequency.

\begin{figure}[!ht]
  \centering
  \includegraphics[width=.50\columnwidth]{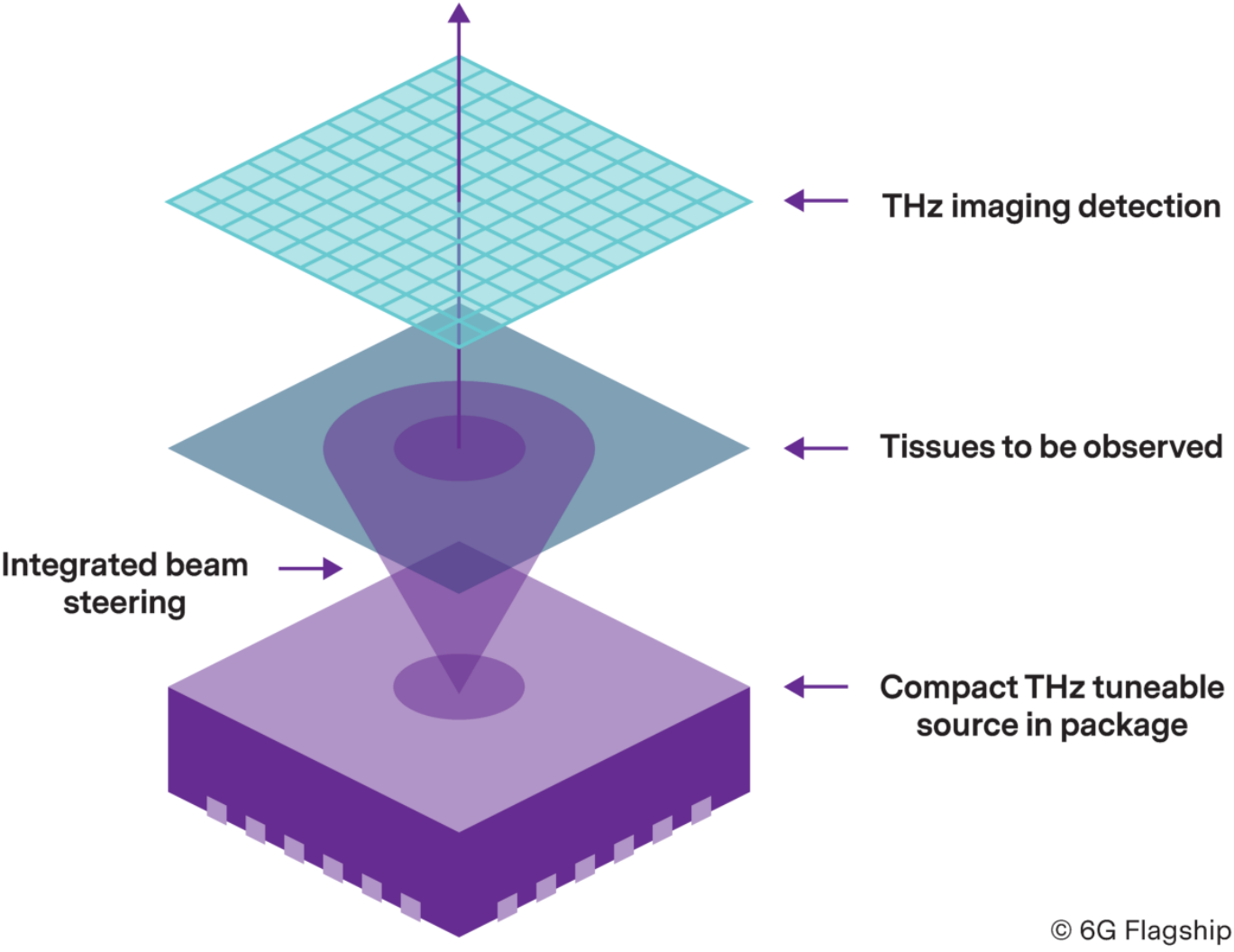}
  \caption{Direct THz imaging.}
  \label{fig:img1}
\end{figure}
\begin{figure}[!ht]
  \centering
  \includegraphics[width=.67\columnwidth]{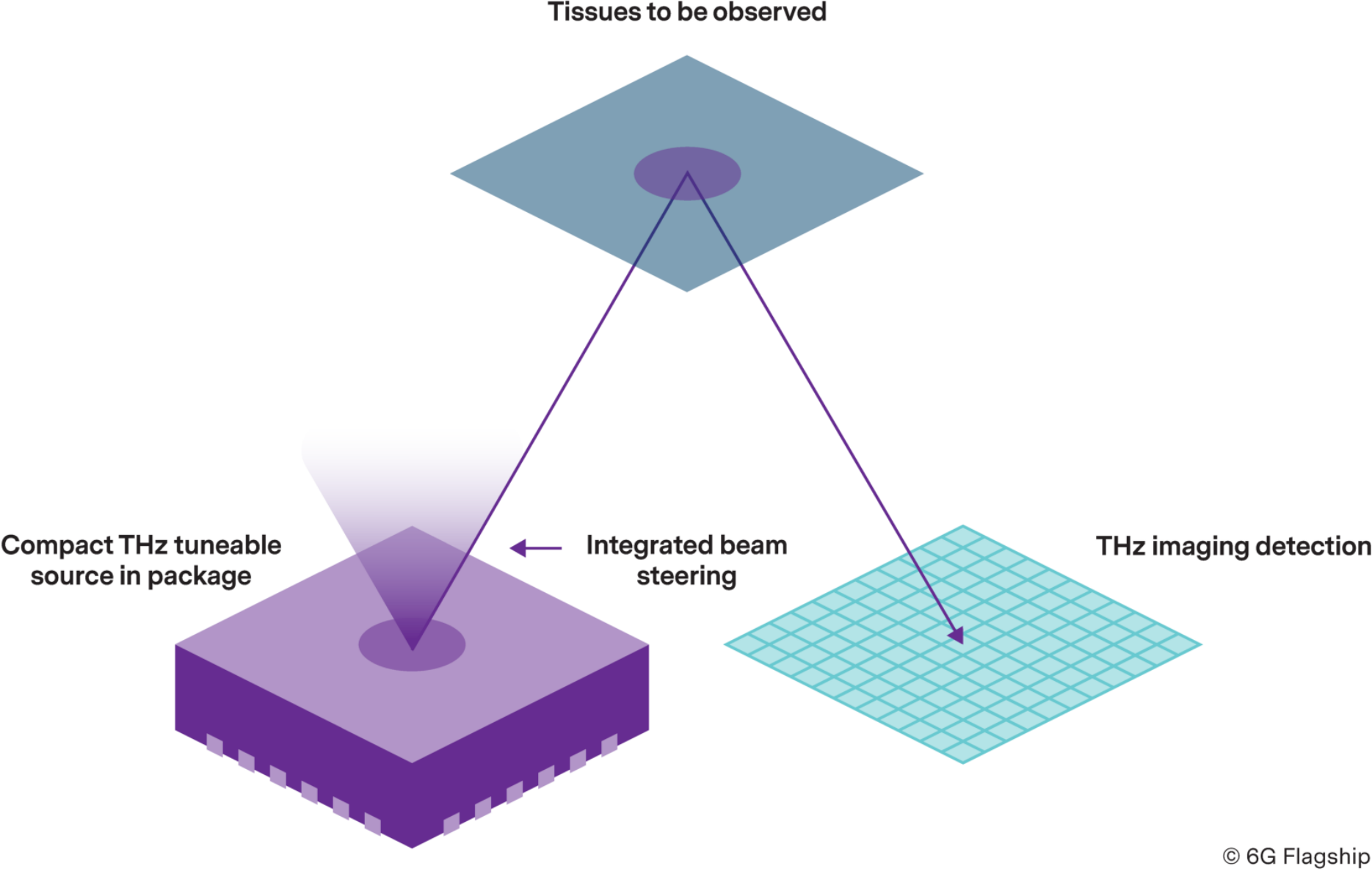}
  \caption{Reflected THz imaging.}
  \label{fig:img2}
\end{figure}
\begin{figure}[!ht]
  \centering
  \includegraphics[width=.53\columnwidth]{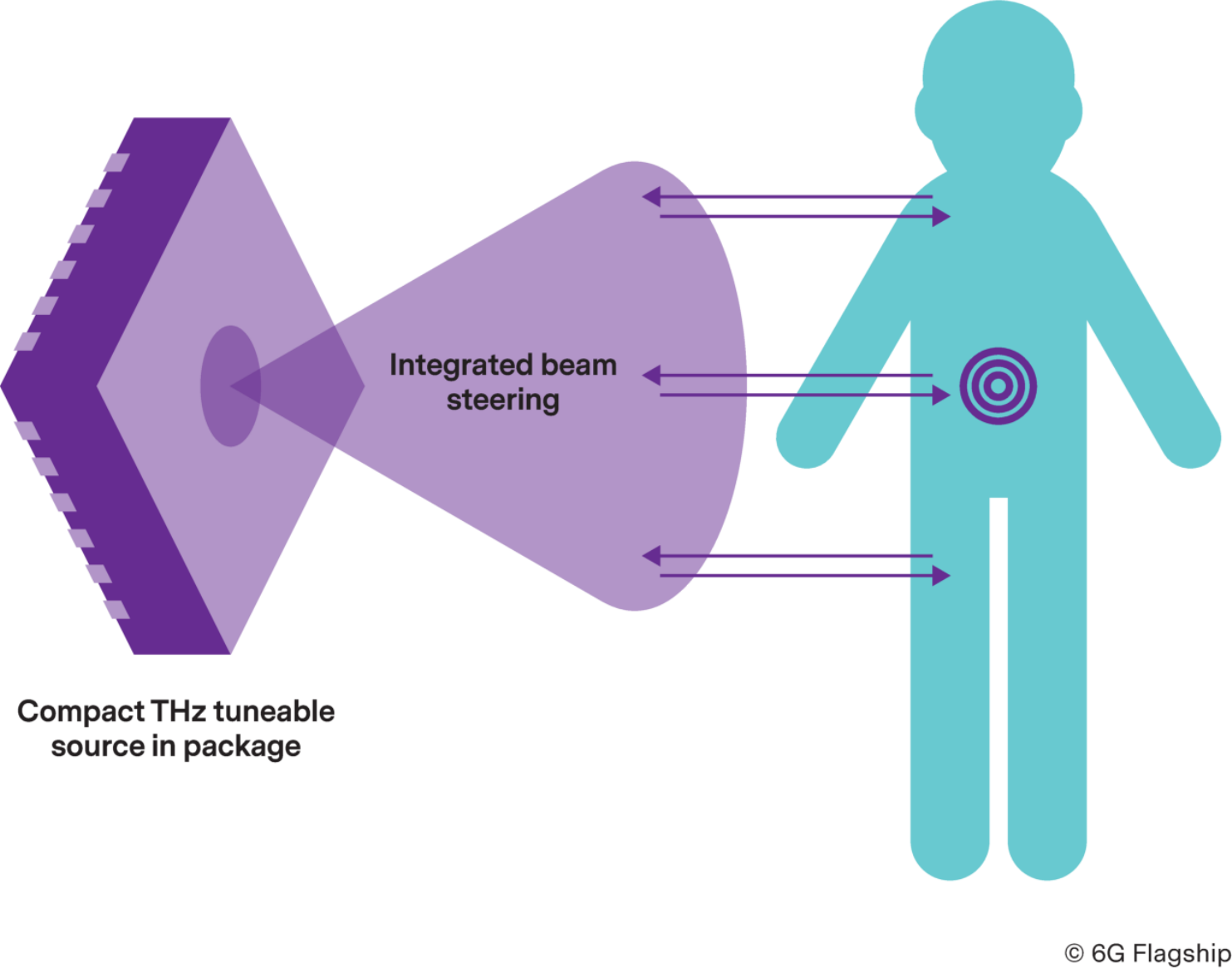}
  \caption{Radar application.}
  \label{fig:img3}
\end{figure}

\subsection{Simultaneous localization and mapping}

In \ac{SLAM}, mobile devices are considered as sensors, with time-varying states (position, pose, as well as their derivatives), and landmarks (object), with fixed or slowly changing states. Both sensor states and landmark states are a priori unknown. A sensor moves through the environment and collects measurements in its local frame of reference related to the landmarks. These measurements may be from \ac{6G} radar-like signals originating from the mobile device or from fixed infrastructure. The sensor has an associated mobility model, describing the evolution of the sensor state statistically as it moves. \ac{SLAM} algorithms aim to recover estimates of the sensor state (including the entire sensor state trajectory) as well as state estimates of the landmarks. 
\ac{SLAM} methods differ in their measurement models, as well as the way they process (online or in a batch) and represent state statistics (e.g., parametric distributions or particles). The most common batch-based method is GraphSLAM \cite{grisetti2010tutorial}, which constructs a factor graph, with vertices that correspond to the poses of the sensor at different points in time and edges that represent measurements. A global optimization is then run to recover the sensor's poses and localize the sensor within the map. It should be noted that beside \ac{SLAM}, the factor graphs running message passing algorithm with the use of various \ac{EM} properties such as \ac{AoA}, \ac{ToA}, \ac{TDoA}, \ac{RSS} and \ac{DRSS} for localization, as well as \ac{EKF} for tracking purpose, have been employed in the past two decades \cite{mcheng2020}. The online methods vary from simple \ac{EKF}-\ac{SLAM} to sophisticated methods based on random finite set theory \cite{mullane2011random}. All these methods must account for unknown data associations between the measurements and landmarks, leading to a high computational complexity.

Performing \ac{SLAM} using \ac{6G} radio signals instead of laser or camera measurements is more challenging than conventional \ac{SLAM} systems because the range and angle measurements are much less accurate and might be subjected to strong outliers because of the non-ideal antenna radiation pattern (sidelobes) and multipath. Therefore, ad hoc measurement models tailored to radio propagation characteristics need to be developed. In particular, the measurements are expected to be rich, comprising angle and delay information, while at the same time the state model is likely to be complex, treating a sensor not as a point with a \ac{3D} location, but also with an orientation in \ac{3D} (with a specific pitch, roll, and yaw angle) \cite{vernaza2006rao}.
%
%
%
%
%
%
%
%
%
%
Moreover, the higher resolution of measurements leads to a fine-grained view of the landmarks in the environment. Hence, refined state models must be considered, e.g., including the \ac{EM} properties of objects, their size and orientation, as is currently done in radar \cite{granstrom2016extended}. 
More specifically, leveraging the geometry side information captured by THz imaging \cite{rappaport-ieee-access19}, the gap between \ac{LIDAR}-based and radio-based \ac{SLAM} will be reduced by opening new appealing perspectives such as the possibility to construct and update automatically accurate maps of indoor environments taking advantage of the pervasiveness of end-user devices, including smartphones and smart glasses, as well as other \ac{XR}-tailored devices in the \ac{6G} era. In fact, user devices could act as \emph{personal radar} devices that scan the surrounding environment obtained from the radio signal range and angle measurements. These raw ``radio images'' measurements can be shared according to the crowd-sensing paradigm and used as input for \ac{SLAM} algorithms. The automatic generation and update of indoor maps will enable new services such as infrastructure-less and map-less localization \cite{GuidiGuerra16}. \ac{6G} \ac{SLAM} methods will also find applications in \ac{AR} / \ac{VR} / \ac{MR} and the localization of autonomous vehicles and drones.     

\subsubsection*{Challenges} The main challenges in \ac{6G} \ac{SLAM} will be the development of novel models for the landmark state and the derivation of powerful, but low complexity algorithms that can run on mobile devices, possibly supported by distributed \ac{MEC}. Due to the extremely high data rates, hardware impairments are expected to be a major limitation. On the other hand, as \ac{SLAM} deals with the movement of objects on slow time scales, e.g., m/s and rad/s, there is plenty of time during which the environment is practically frozen to process and collect the measurements.  

\subsection{Passive sensing using transmitters of opportunity}
\label{sec:passive_sensing}
Passive sensing (sometimes referred to as passive radar or passive coherent location) is an emerging technology by which the energy reflected by static or moving reflectors is received and processed by a receiver device to extract information about the objects. It is passive in the sense that no signal is transmitted for this purpose: the receiver opportunistically uses radio waves (usually wireless communication signals) transmitted for other uses such as cellular, Wi-Fi and TV or radio broadcasting signals. It can be viewed as a bi-static radar where the transmitter is not ``controlled by'' or ``synchronized with'' the receiver. The receiver typically uses prior knowledge of the physical layer and frame structure of the wireless standard in use to extract information about the environment. Although not strictly necessary, it is easier for the processing receiver if it uses a separate antenna channel to receive the \ac{LOS} signal from a wireless transmitter. Signals from stationary transmitters (\ac{BS}, \ac{AP} or broadcast antennas) are preferred since echoes from static reflectors then have zero Doppler, although static clutter removal may be challenging \cite{Storrer2020}. Passive sensing is also non-cooperative in nature since detected objects and targets do not communicate with the receiver, nor do they carry special reflectors or transponders to help the receiver.
For moving targets, passive sensing allows several interesting applications such as vital sign monitoring, fall detection, presence detection, intruder detection, human activity recognition and localization and more \cite{Savazzi2019}. In addition, passive sensing could potentially help to create maps of the static environment. Passive sensing has also been reported in through-the-wall detection/imaging.

We foresee an increased interest in and deployment of passive sensing, supported by the progress in signal processing, compressive sensing and machine learning. Going in this direction, the IEEE \(802.11\) Working Group created the ``WLAN Sensing Study Group'' that is studying the creation of a Task Group to draft the \(802.11\) standard for WLAN sensing \cite{IEEE802_SENS}. Very advanced examples include the use of Wi-Fi signals to to carry out through-the-wall imaging \cite{Tan2016}, the use of digital video broadcasting signals to carry out \ac{2D} passive radar imaging with inverse synthetic aperture radar (ISAR) techniques \cite{Garry2019_1} and \cite{Garry2019_2} and the use of \ac{5G} signals to locate road users \cite{Thoma2019}.
Signal processing for passive sensing borrows much from radar signal processing. A major difference is in the waveform design since chirp-based waveforms are not common in wireless communications, and the wireless modulation is typically not optimized for the (range-Doppler) ambiguity function. Therefore, the matched filter processing is usually different. Building range-Doppler maps and range-Doppler-angle radar data cubes is the ultimate goal of passive sensing, and this goal is often not achieved with the same quality and \ac{SNR} as for conventional active radars. The whole arsenal of \ac{ML} methods (see Section \ref{sec:ai_ml}), including hand-crafted feature extraction, deep learning and even spiking neural networks, could be adapted to passive sensing.

\subsubsection*{Challenges}
Because of its opportunistic nature and the impossibility to control the transmission, passive sensing will almost always achieve a lower degree of performance than a conventional active radar. The challenges that must be addressed to improve the performance of passive sensing in \ac{6G} networks are numerous: improving the quality (\ac{SNR}, multipath-free) of the reference signal; maintaining the coherence-on-receive over long observation times towards fine Doppler resolution; removing the static clutter and coping with multipath reflections which create ghosts; accommodating the limited bandwidth, and hence the range resolution of the wireless signals (e.g. a \(20\)~MHz signal translates into a range resolution of \(15\)~m); in the spatial dimension, improving the angular resolution and coping with the non-uniform illumination due to the environment and transmitter beamforming. The implementation of low-cost, low-power terminals involves several challenges, such as carefully balancing the use of data-aided (i.e., preamble) and non-data aided (i.e., communication payload) signals and reusing existing \ac{RF} and digital hardware.

\subsection{Active sensing with radar and communications convergence}

The use of radar was for a very long time limited to applications in the military, aviation, law enforcement and meteorology using large and expensive antennas and equipment. However, more recently, advances in semiconductor technology, antenna design and signal processing \cite{Hasch2012, Bilik2019} have made low-cost integrated radars readily available. On account of this, radar usage has also exploded in the consumer market, particularly in the automotive industry, where features such as adaptive cruise control, lane change assistance and cross-traffic alerts rely on radar sensing. Beyond this, new applications are emerging, such as intruder detection, gesture recognition, and heart rate and respiration rate monitoring, among many others.
Radar detects the range, angle and velocity of objects based on the propagation of 
\ac{EM} waves, which are reflected by these objects, and, as such, rely on the same physical phenomena as wireless communications. In spite of this, current radar systems use dedicated chipsets and antennas, and have reserved frequency bands, for instance, in the \ac{UWB}, \(24\)~GHz and \(76\)-\(81\)~GHz bands. However, wireless communication technologies are converging towards the requirements of radar systems, with the increasing usage of directive antenna patterns through beamforming and transmission at higher frequencies with wider bands, which are required for angle and range resolution. The rise of new applications and market demand, together with technological developments both in the radar and in the wireless communications industry, make the integration of both technologies in a single system desirable and feasible in the near future.

\begin{figure}[h]
  \centering
  \includegraphics[width=.75\columnwidth]{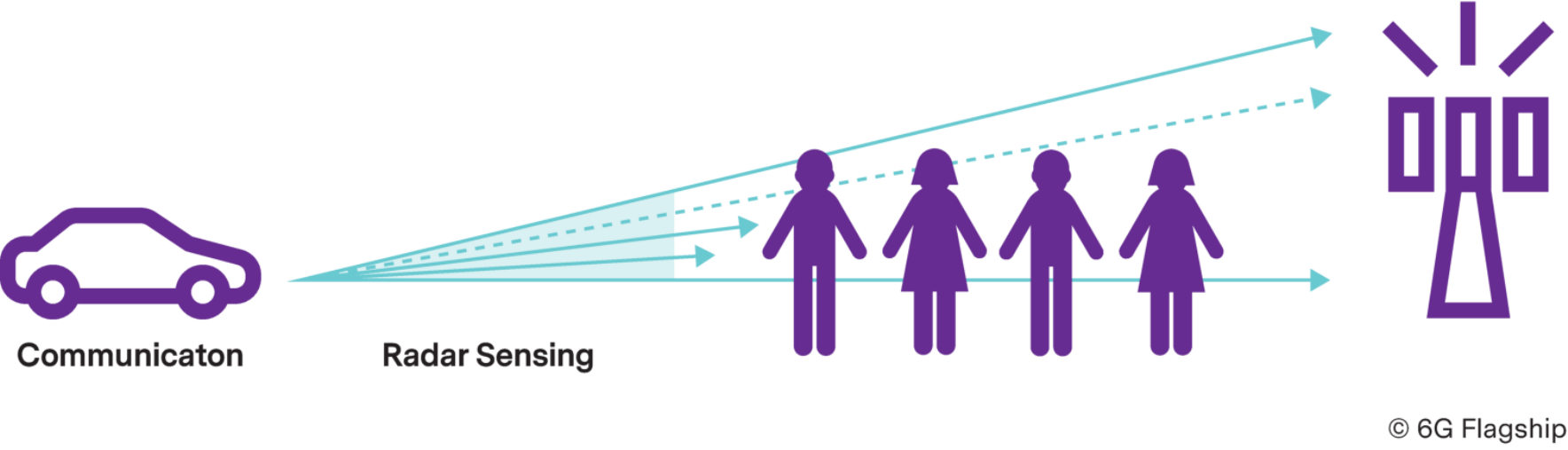}
  \caption{Interplay between communications and radar.}
  \label{fig:radcomm}
\end{figure}

There are several ways in which radar and communications may operate in the same spectrum: coexistence, cooperation, and co-design. Coexistence uses cognitive-radio techniques and beamforming is used to avoid interference in time and space domains \cite{Zheng2019}. This is a good approach when large-scale radar systems, such as aeronautical or meteorological radars are considered, but not for more dynamic scenarios and applications, as in the automotive branch or in the consumer market. In such cases, cooperation is an option, whereby the radar and communication sub-systems operate largely independently, but exchange information to support each other, in particular for interference mitigation. For instance, in-band communications may also help radar detection, in that it allows the exchange of information between different sensors, and, on the other hand, radar may help communications, for instance, by gathering information about obstacles and reflectors, that may aid beam tracking.
As a third alternative, co-design is a more promising approach, in which a single waveform is employed both for radar and communications. This is likely to be the path taken in \ac{6G} systems, and it brings about advantages and opportunities, such as more efficient and flexible use of spectrum for different purposes, as well as lower costs through hardware reuse and integration and lower energy consumption than in two separate systems. The concept of a joint radar and communications system is depicted in Fig.~\ref{fig:radcomm}.

As already mentioned in Section \ref{sec:imaging} and Section \ref{sec:passive_sensing}, \ac{6G} systems will probably rely on both passive and active radar principles. 
In active radar, as we can leverage the knowledge of the transmitted signal as a reference, more precise and accurate results can be expected, but some hardware and signal-processing challenges may arise, particularly to deal with a strong self-interference at the radar receiver. Both passive and active radars are likely to be employed simultaneously, and the information provided by both can be combined by means of sensor fusion to provide a rich and accurate mapping of the environment. In any case, with new applications and requirements in mind, the \ac{6G} waveform will need to be designed so that the following features are enabled: (i) good radar performance in terms of range and resolution in \ac{4D} (range, velocity, azimuth and elevation angle); (ii) good communication performance in terms of spectral and power efficiency; (iii) flexibility between communications and sensing needs as well as for different radar needs (e.g., short-range/long-range and different resolutions).

\subsubsection*{Challenges}
While communicating and sensing with the same waveform promises many advantages, there are some significant challenges to be solved.
The waveform design must satisfy trade-offs between the communications and radar requirements listed above. Improved radar detection and estimation algorithms are needed for wireless communications scenarios, which are characterized by high mobility, multiple targets and clutter. Techniques employing compressed sensing and machine learning may be required. Additionally, suitable hardware architecture is needed, including antenna design, duplexing, analog-to-digital conversion, mixers, accounting for different radar and communications receivers.
Interference is likely to be a difficulty. Self-interference is a significant issue in active radars. In-band full-duplexing is de-facto needed, and the antenna, waveform and signal processing have to be designed to minimize its impact as much as possible. The increased usage of radar for different applications will also result in higher interference levels between different users. Interference management and cancellation techniques will be needed, although the large difference in transmission powers needed for communication and reliable radar detection will make this challenging.
However, the challenges are not all related to signal processing and the physical layer. New \ac{MAC} protocols and \ac{RRM} algorithms will be needed to allocate the radio resources according to the needs of different radar and communication services. A better understanding of the trade-offs between radar and communications \acp{KPI} may be necessary.


\subsection{Channel charting}\label{sec:channelcharting}

\begin{figure}[!h]
  \centering
  \includegraphics[height=.875\textheight]{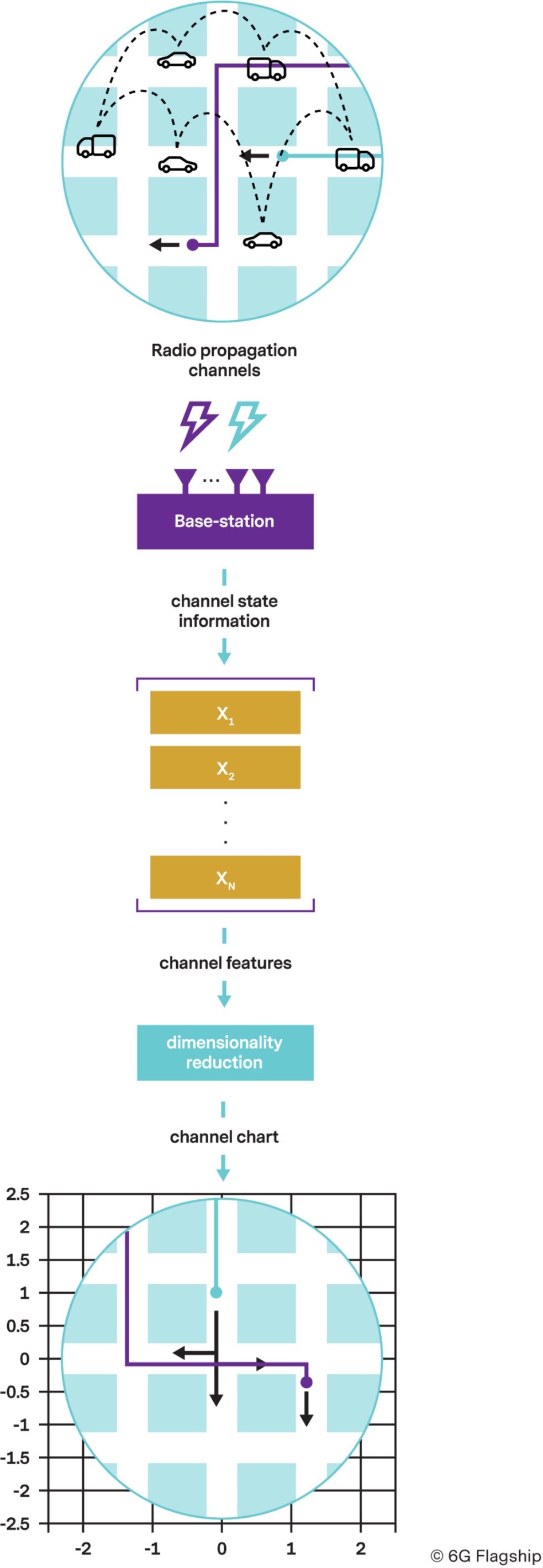}
  \caption{Channel charting uses dimensionality reduction techniques to generate a pseudo-map of the wireless propagation channel, and associates the user \ac{CSI} to a pseudo-location which can be tracked consistently across the users and over time.}
  \label{fig:channelcharting}
\end{figure}

Channel charting \cite{Studer_etal_charting} is a baseband processing approach involving applying classical unsupervised dimensionality reduction methods from machine learning to \ac{CSI}. Based on a large dataset of \ac{CSI} samples acquired in a given environment, it creates a virtual map (known as a \emph{chart}) in an unsupervised manner, on which the users can be located and tracked. Due to the unsupervised nature of the approach, channel charting is applicable to situations where there is not enough information available about the geometric properties of the scattering environment to build a faithful geometric model of the user's environment and location. Despite the fact that a position on the chart cannot be readily mapped to a user location, it provides a real-time pseudo-location within the cell, which is consistent across the users and over time (Fig.~\ref{fig:channelcharting}), without the need for expensive dedicated measurement campaigns to obtain a labeled \ac{CSI} dataset as required by fingerprinting methods, for example. The tracking of the pseudo-location on the chart can be used to enhance numerous network functionalities, e.g., for predictive \ac{RRM} and rate adaptation, handover between cells, \ac{mmWave} beam association and tracking and  \ac{UE} pairing or grouping in \ac{D2D} scenarios.

Although it does not constitute a universal replacement for a true position in location-based applications, the use of a pseudo-location has certain benefits. 
First, the unsupervised nature of channel charting allows some form of self-configuration that does not involve any prior information (such as an area map, or knowledge about the geometry of the surrounding buildings); this is a handy feature for the deployment of temporary or emergency networks.
Pseudo-location can also be seen as a privacy protection feature, allowing applications such as contact tracing to be implemented without ever requiring the actual user position to be estimated.

\subsubsection*{Challenges}
Channel charting has been developed initially within the massive \ac{MIMO} paradigm, where scattering is rich and prominent \ac{RF} propagation features are expected to be stable (or evolve slowly) over time. 
Applying this approach to higher frequency bands (where line-of-sight propagation dominates) will be likely require extending it to jointly process signals from multiple transmission points.
Other open questions associated with charting stem directly from the machine-learning roots of channel charting, such as the ability to implement life-long learning, or optimal feature design for \ac{CSI} signals.
    
\subsection{Context-aware localization systems}

\ac{6G} should allow the phenomenological interpretations of ``context'', i.e., it should allow interpretation of the situation of entities through action. Physical, cultural and organizational contexts shape actions and create the conditions for humans to interpret and understand their actions. This is done not only by the measurable characterization of entities such as person, place or object but also by combining this with the understanding of the activity and temporal aspect of behavior. Context-awareness is equally important in detecting the contextual factors of devices, and humans \cite{dourish2001seeking}.

\ac{6G} applications will be able to benefit from context-awareness in several ways. Applications will consume less energy because an application knows when it is ideal to communicate. Context-awareness allows intelligent prediction of data transmission, which can allow better throughput on demand. Context-awareness can be used to move personalization algorithms and sensor data intelligently into parts of the network where storage and computation are fast and feasible, which would allow the hyper-personalization of services. Intelligent storage and distributed processing capabilities will allow the fusion of sensor data for detecting trends and deviations, which are essential for example in healthcare scenarios \cite{dourish2001seeking, dey2001conceptual}.
Context-awareness also allows multi-modal localization, where mobile devices can switch between different channels and communication technologies depending on the current location or context. Multi-modality enables devices to reduce their power consumption, increase the quality of service, and select local or private technologies versus national or public technologies. As a consequence of these differences in communication modalities, different localization methodologies must be implemented as well, based on context-awareness \cite{nust2019visualizing}. 
\ac{6G} can also boost the following specific trends in context-awareness. First of all, it will be able to detect trends and deviations, i.e., understand the temporary aspects of context through storage of context history, and processing of temporal context data. This is needed, for example, in hyper-personalization, where real-time deviations of a person’s status are detected. Secondly, advanced temporal context detection algorithms have special security concerns as they combine highly personal data (such as physiological measures of humans) with open, public data (such as weather conditions). Storing and processing this data in different parts of the infrastructure requires intelligent security solutions. Finally, distributed context-awareness requires a high level of standardization of context parameters. Without standardized interpretation rules for context parameters, it will be impossible to build connectivity infrastructures that can translate the context into assets for applications.

\subsubsection*{Challenges}
Challenges in implementing and adoption of these features include, the standardization of context parameters and operating principles for fair data economy facilitating flow and combinations of different data sources governed by a heterogeneous ecosystem.

\subsection{Security, privacy and trust for localization systems}
It is expected that more and more stakeholders will be involved in the \ac{6G} positioning chain, i.e., virtual network operators, and that there will be an increased number of location-based service providers due to the developing new applications and services. There will also be multi-user devices supporting cooperative and opportunistic services. All such stakeholders in the positioning chain will be able to benefit from the availability of new security, privacy, and trust solutions for localization systems. For example, secure location information will be able to be increasingly used as a security parameter for digital interactions in more general contexts, e.g., automated driving, health monitoring, social media, surveillance systems, etc. 
In order to reduce the power consumption of user devices and to enable massive location-based \ac{IoT} connectivity, it is likely that network- and cloud-based localization solutions will become much more widespread than device- and edge-based localization solutions. This raises the questions of trust in the network- and cloud-service providers, as well as in the location-lased service providers. 

\subsubsection*{Challenges}
Service providers’ vulnerabilities could be exploited by possible attackers  to extract user’s location patterns and to misuse the information for identity thefts, burglaries, toll avoidance, stalking, etc. \cite{LAA+2018,LRLInsideGNSS2017}.
Privacy concerns due to the future use of THz communications on hand-held devices and wearables have been already raised, e.g., in \cite{sarieddeen2019next}. An attacker or a malicious device could conduct THz-based remote sensing and see-through imaging that could be privacy-invasive. Radar-like localization solutions are also emerging base on \ac{4G} and \ac{5G} signals and advances in full-duplex communications \cite{BAF+2019,GKG+2019,8805161}, and it is likely to be expanded beyond \ac{5G} too. High-resolution imaging combined with machine learning techniques could be invasive of privacy; for example, \ac{RFF} \cite{SNY+2020,BGB+2020} can identify user devices even in the absence of transmitting a device identity.

\section{Summary and research questions}
\label{sec:questions}
In this contribution, we identified not only several key technological enablers toward \ac{6G} localization and sensing systems but also novel applications and service opportunities. These enablers correspond to the new \ac{RF} spectrum at the high-frequency range, especially above \(100\)~GHz and beyond; the introduction of intelligent reflective surfaces which allow network operators to shape and control the \ac{EM} response of the environment; advanced beam-space processing to track users and objects, as well as map the environment; the pervasive use of artificial intelligence leveraging the unprecedented availability of data and computing resources to tackle fundamental problems in wireless systems; and advances in signal processing to support novel convergent communication and radar applications. In turn, these enablers will lead to new opportunities in the \ac{6G} era. The opportunities highlighted in this white paper were imaging for biomedical and security applications; applications of simultaneous localization and mapping to automatically construct maps of complex indoor environments; passive sensing of people and objects; using location information as a big data source, guiding and predicting the human-digital ecosystem; the coexistence and cooperation between sensing, localization and communication, leading to one device with this three-fold functionality; and finally, the use of location information to boost security and trust in \ac{6G} connectivity solutions. 

A joint effort across a large number of scientific disciplines will be required to achieve these goals, much more than the effort required in any previous mobile communication generation. In order to realize these opportunities, fundamental research questions need to be properly addressed. Below are a selection of key questions to be addressed:
\begin{enumerate}
\item How can high-accuracy cm-level positioning and high-resolution \ac{3D} sensing/imaging be achieved by taking advantage of the \ac{6G} key technology enablers?

\item How can novel waveform designs be devised so as to enable convergent communication, localization and sensing systems which efficiently share resources in time, frequency and space domains?

\item How can energy-efficient high-accuracy localization and high-resolution sensing/imaging solutions be devised while operating at a high frequency range and supporting mobility of the communicating objects?

\item How can real-time energy-efficient \ac{AI}/\ac{ML} techniques be developed to achieve high-accuracy localization and high-resolution sensing solutions, while leveraging the unprecedented availability of data and computing resources?

\item How can the quality and accuracy gap between passive sensing and active sensing be bridged?
\end{enumerate}

\newpage

{\setstretch{1.1}
\section*{List of acronyms}
\begin{multicols}{2}
\begin{acronym}[ACRONYM]
\acro{2D}{two-dimensional}
\acro{3D}{three-dimensional}
\acro{4D}{four-dimensional}
\acro{4G}{fourth generation}
\acro{5G}{fifth generation}
\acro{6G}{sixth generation}
\acro{ADAS}{advanced driver assistance systems} 
\acro{AI}{artificial intelligence}
\acro{AoA}{angle of arrival}
\acro{AoD}{angle of departure}
\acro{AP}{access point}
\acro{AR}{augmented reality}
\acro{BS}{base station}
\acro{CMOS}{complementary metal–oxide–semiconductor}
\acro{CSI}{channel state information}
\acro{CUPS}{control and user plane separation}
\acro{D2D}{device to device}
\acro{DRSS}{differential received signal strength}
\acro{E911}{Enhanced 911}
\acro{EKF}{extended Kalman filter}
\acro{EM}{electromagnetic}
\acro{FCC}{Federal Communications Commission}
\acro{FD-SOI}{fully depleted silicon on insulator}
\acro{GaAs}{Gallium Arsenide}   
\acro{GaN}{Gallium Nitride}      
\acro{GNSS}{Global Navigation Satellite System}
\acro{IIoT}{Industrial IoT}
\acro{InP}{Indium Phosphide}
\acro{IoT}{Internet of Things}
\acro{IR}{infrared} 
\acro{IRS}{intelligent reflective surface}
\acro{KPI}{key performance indicator}
\acro{LIDAR}{light detection and ranging}
\acro{LO}{local oscillator}
\acro{LOS}{line of sight}
\acro{MAC}{medium access control}
\acro{MAS}{multi-antenna system}
\acro{MEC}{mobile edge computing}
\acro{MEMS}{microelectromechanical systems}
\acro{MIMO}{multiple input multiple output}
\acro{ML}{machine learning}
\acro{mmWave}{millimeter wave}
\acro{MR}{mixed reality}
\acro{muWave}[\(\mu\)mWave]{micrometer wave}
\acro{NFmin}{Noise Figure Min}
\acro{NLOS}{non line of sight}
\acro{NR}{new radio}
\acro{RF}{radio frequency}
\acro{RFF}{radio frequency fingerprinting}
\acro{RRM}{radio resource management}
\acro{RSS}{received signal strength}
\acro{SiGe}{Silicon Germanium}
\acro{SIP}{system in package}
\acro{SLAM}{simultaneous localization and mapping}
\acro{SNR}{signal to noise ratio}
\acro{TDoA}{time difference of arrival}
\acro{TDS}{time domain spectroscopy}
\acro{ToA}{time of arrival}
\acro{UE}{user equipment}
\acro{UWB}{ultra wideband}
\acro{VR}{virtual reality}
\acro{XR}{cross reality}
\end{acronym}
\end{multicols}
}

\bibliographystyle{IEEEtran}
\bibliography{IEEEabrv,whitepaper_levi20}

\end{document}